\documentclass[a4paper,11pt]{article}
\usepackage{ILD}
\RequirePackage{rotating}
\usepackage{subcaption}
\usepackage{float}
\usepackage[affil-it]{authblk}
\usepackage{booktabs}
\usepackage{amssymb}
\usepackage{siunitx}

\usepackage[backend=biber,style=numeric-comp,sorting=none,mcite=true,
doi=false,subentry]{biblatex}
\usepackage{hyperref}
\usepackage{pgfpages}
\usepackage{fancybox,graphicx}
\usepackage{graphbox}
\usepackage{amsfonts}
\usepackage{amsmath}
\usepackage{relsize}
\usepackage{slashed}
\usepackage{wrapfig,rotating}
\usepackage{tikz}
\usepackage{pgfpages}
\usepackage{fancybox,graphicx}
\usepackage{epstopdf}

\AppendGraphicsExtensions{.eps.gz}
\usepackage{times}
\usepackage[T1]{fontenc}

\usepackage{lineno}

\newcommand\snowmass{\begin{center}\rule[-0.2in]{\hsize}{0.01in}\\\rule{\hsize}{0.01in}\\
\vskip 0.1in Submitted to the  Proceedings of the US Community Study\\ 
on the Future of Particle Physics (Snowmass 2021)\\ 
\rule{\hsize}{0.01in}\\\rule[+0.2in]{\hsize}{0.01in} \end{center}}

\newcommand{\alc}[2]{\multicolumn{1}{#1}{#2}}
\newcommand{\tento}[1]{\mbox{$\cdot 10^{#1}$}}
\title{Evaluating the ILC SUSY reach in the most challenging scenario: \\
  $\widetilde{\tau}$ NLSP, low $\Delta M$ , lowest cross-section.}
\ildproc{phys}{2022}{005}
\date{\today}
\addauthor{M.T. N{\'u}{\~n}ez Pardo de Vera}
{\institute{1}}
\addauthor{M. Berggren}
{\institute{1}}
\addauthor{J. List}
{\institute{1}}

\addinstitute{1}{  Deutsches Elektronen-Synchrotron DESY,\\
                      Notkestr. 85, 22607 Hamburg, Germany}

\abstract{
The direct pair-production of the superpartner of the $\tau$-lepton, the $\widetilde{\tau}$,  is one
of the most interesting channels to search for SUSY in. First of all, the $\widetilde{\tau}$ is
likely to be the lightest of the scalar leptons. Secondly the
signature of $\widetilde{\tau}$ pair production is one of the experimentally most difficult
ones, thereby constituting the ``worst'' possible scenario for SUSY searches.
The current model-independent $\widetilde{\tau}$ limits comes from analyses performed at
LEP but they suffer from the limited energy of this facility. Limits obtained at the LHC do
extend to higher masses, but they are only valid under strong assumptions.
The International Linear Collider, the ILC, is a future electron-positron collider, foreseen
to operate initially at a centre-of-mass energy of 250 GeV,
then to be upgraded 
to 500 GeV, and possibly to 1 TeV at a later stage.
ILC will be a powerful facility for SUSY
searches. 
The capability of the ILC for determining exclusion/discovery limits
for the $\widetilde{\tau}$ in a model-independent way is shown in this paper, together
with an overview of the current state-of-the-art.
Results of the last studies of $\widetilde{\tau}$ pair-production at the ILC are presented,
showing the improvements with respect to previous results.
A detailed study of the ``worst'' scenario for $\widetilde{\tau}$ exclusion/discovery, taking
into account the effect of the $\widetilde{\tau}$ mixing on $\widetilde{\tau}$ production cross-section
and detection efficiency, is presented.
The study also includes an analysis of the effect of the overlay particles in the $\widetilde{\tau}$ searches.
The conclusion is that both the exclusion and discovery reaches for this ``worst'' case would extend to
only a few GeV below the kinematic limit at the ILC. Also scenarios with the  $\widetilde{\tau}$
and the Lightest SUSY Particle (the LSP) quite close in mass can be discovered or excluded
at most $\widetilde{\tau}$ masses.
The studies were done using detailed detector simulation of the ILD concept at the ILC.
For signal, the fast detector simulation {\tt SGV} was used, while the full Geant4 based {\tt DDSim} was
used for the standard model backgrounds.

\snowmass
}
\begin{document}
\titlepage

\section{Introduction\label{sec:intro}}
  Supersymmetry (SUSY)
     \cite{Martin:1997ns,Wess:1974tw,Nilles:1983ge,Haber:1984rc,Barbieri:1982eh}
    is one of the most promising candidates for new physics. It could explain or
  at least give some hint at solutions to current problems of the Standard Model (SM), such as the gauge hierarchy
  problem, the nature of Dark Matter (DM) or the possible theory-experiment discrepancy of the muon magnetic moment. 
  SUSY is a symmetry of
  spacetime relating fermions and bosons. For every SM particle it introduces a superpartner with the
  same quantum numbers except for the spin. The spin differs by half a unit from the value of its SM partner.
  A new parity, R-parity, is commonly introduced in SUSY, which has a crucial impact in SUSY phenomenology.
  R-parity takes an even value for SM particles and odd value for the SUSY ones.
  Multiplicative R-parity conservation~\footnote{The introduction and conservation of
  this symmetry is inspired by flavour physics constraints since the most general SUSY Lagrangian induces
  flavour-changing neutral interactions that are avoided imposing R-parity conservation.}, assumed in
  most of the SUSY models, implies that the SUSY particles are always created in
  pairs and that the lightest SUSY particle (the LSP) is stable and,  when cosmological constraints are taken into account,
  also neutral.
  An important point in this kind of studies is the fundamental SUSY principle stating that 
  the couplings of particles and sparticles are related by symmetry.
  This allows to know the cross sections
  for SUSY pair production solely from the knowledge of initial centre-of-mass energy of the collider 
  and the masses of the involved SUSY
  particles.

\section{SUSY searches\label{sec:susysearch}}
  Considerable efforts have been and are being devoted to the search of SUSY at different facilities.
  Searches at hadron colliders, such as the LHC, are mainly sensitive to the production of coloured sparticles,
  i.e. the gluino and the squarks. They are most probably the heaviest ones. 
  The search of the lighter colour-neutral SUSY states, such as sleptons, charginos or neutralinos, at
  hadron colliders is challenged by the much smaller cross sections,
  and high backgrounds. Mass limits have been obtained at the LHC, but they
  are only valid if many constraints on the model parameters are fulfilled.
  Lepton colliders, like LEP, have a higher sensitivity to the production of colour-neutral SUSY states, but
  the searches up to now were limited by the beam energy. Limits computed at these facilities are however
  valid for any value of the model parameters not shown in the exclusion plots.
  The future International Linear Collider (ILC)~\cite{Behnke:2013xla}, an electron-positron collider operating at centre-of-mass
  energies of $250-500$\,GeV and with upgrade capability to $1$\,TeV, is seen as an ideal environment for SUSY studies.
  SUSY searches at the ILC would profit from the high electron and positron beam polarisations,
  $\pm$80$\%$$\mp$30$\%$ \footnote{We introduce the following notation for beam polarisations,
$\mathcal{P}\equiv( \mathcal{P}_{e^-}, \mathcal{P}_{e^+} )$,
and define the pure beam polarisations as
$\mathcal{P}_{LR}\equiv(-1, +1)$ and $\mathcal{P}_{RL}\equiv(+1, -1)$.
The nominal beam polarisations for the ILC are defined as
$\mathcal{P}_{-+}\equiv(-0.8, +0.3)$ and $\mathcal{P}_{+-}\equiv(+0.8, -0.3)$.}
 in this study, 
  a well defined initial state (in 4-momentum and spin configuration),
  a clean and reconstructable final state, near absence of pile-up, hermetic detectors (almost
  4$\pi$ coverage) and trigger-less operation, which is a huge advantage for precision measurements and
  unexpected signatures.


\section{Motivation for $\widetilde{\tau}$ searches\label{sec:motivation}}
  For evaluating the power of SUSY searches at future facilities,
  it is beneficial to focus on the lightest particle in the SUSY spectrum that could be accessible. Since
  the cosmological constraints requires a neutral and colourless LSP, the next-to-lightest SUSY
  particle, the NLSP,  would be the first one to be detected. The NLSP can only decay to the LSP and
  the SM partner of the NLSP (or ``virtually'' {\it via} its SM partner, if the LSP-NLSP mass-difference is smaller than the mass
  of the SM partner). This already makes the NLSP production special: heavier states might well decay in
  cascades, and thus have signatures that depend strongly on the model.
  Furthermore, there is only a finite set of sparticles that could be the NLSP, so a systematic search
  for each possible case is feasible.
  This also means that one can a priori estimate which will be the most difficult case, namely the
  NLSP that combines small production cross-section with a difficult experimental signature.
  The $\widetilde{\tau}$ satisfies both these conditions. Therefore, studies  of $\widetilde{\tau}$ production
  might be seen as the way to determine the guaranteed discovery or exclusion reach for SUSY: any other NLSP
  would be easier to find.
  
  The $\widetilde{\tau}$ is the super-partner of the ${\tau}$. Like for any other fermion or sfermion, there
  are two weak hyper-charge states, $\widetilde{\tau}_{R}$ and $\widetilde{\tau}_{L}$. For the fermions the
  chiral symmetry assures that both weak hyper-charge states are degenerate in mass. However this symmetry
  does not apply to sfermions, since they are scalars, rather than fermions. Hence there is no reason to
  expect that  $\widetilde{\tau}_{R}$ and $\widetilde{\tau}_{L}$ would have the same mass.
  Furthermore, mixing between the weak hyper-charge states
  yields the physical states. The strength of the couplings involved in the mixing of states depend on the fermion mass and hence only
  the third generation of the sleptons, $\widetilde{\tau}$, will mix~\footnote{This is also the case for the
  squarks, where the third generation, the stop the and the sbottom, are expected to mix.}. 
  As a consequence of the mixing the lightest
  $\widetilde{\tau}$, $\widetilde{\tau}_{1}$, would most likely  be the lightest slepton, due to the seesaw
  mechanism: the mass of the lightest physical (mixed) state would be smaller than the mass of
  any un-mixed weak hyper-charge state. The cross section of the $\widetilde{\tau}$ also differs from the
  one of the ${\tau}$, not only due to phase-space limitations - the $\widetilde{\tau}$ being more massive
  than the ${\tau}$ - but also due to the mixing.
  In $e^+e^{-}$ colliders, assuming R-parity conservation, the $\widetilde{\tau}$ will be pair-produced, with
  contribution of the s-channel only, via $Z^{0}$/$\gamma$ exchange. The strength of 
  the  $Z^{0}$/$\gamma$ $\widetilde{\tau}$ $\widetilde{\tau}$ coupling depends on 
  the $\widetilde{\tau}$ mixing, reaching its minimum value when
  the coupling $\widetilde{\tau}_{1}$ $\widetilde{\tau}_{1}$ $Z^{0}$ vanishes, which will lead to the worst possible
  scenario in $\widetilde{\tau}$ and, in general, in slepton searches. 
  Another property making the search of  $\widetilde{\tau}$ the worst case, is the fact that its SM partner
  is unstable. It decays before it can be detected, and, as a further complication, some of its decay products are
  undetectable neutrinos.
  This on one hand  makes the identification 
  more difficult than the direct decay to electrons or muons, and on the other hand,
  since the decay products are only partially detectable, that kinematic signatures get blurred.
  The search of a light $\widetilde{\tau}$ is also theoretically motivated: SUSY models with a light
  $\widetilde{\tau}$ can accommodate the observed relic density, by enhancing 
  the $\widetilde{\tau}$-neutralino co-annihilation process~\cite{Ellis:1998}. 

\section{Limits at LEP, LHC and previous ILC studies\label{sec:otherlimits}}
\begin{figure}[htbp]
    \centering
    \subcaptionbox{}{\includegraphics [scale=0.30]{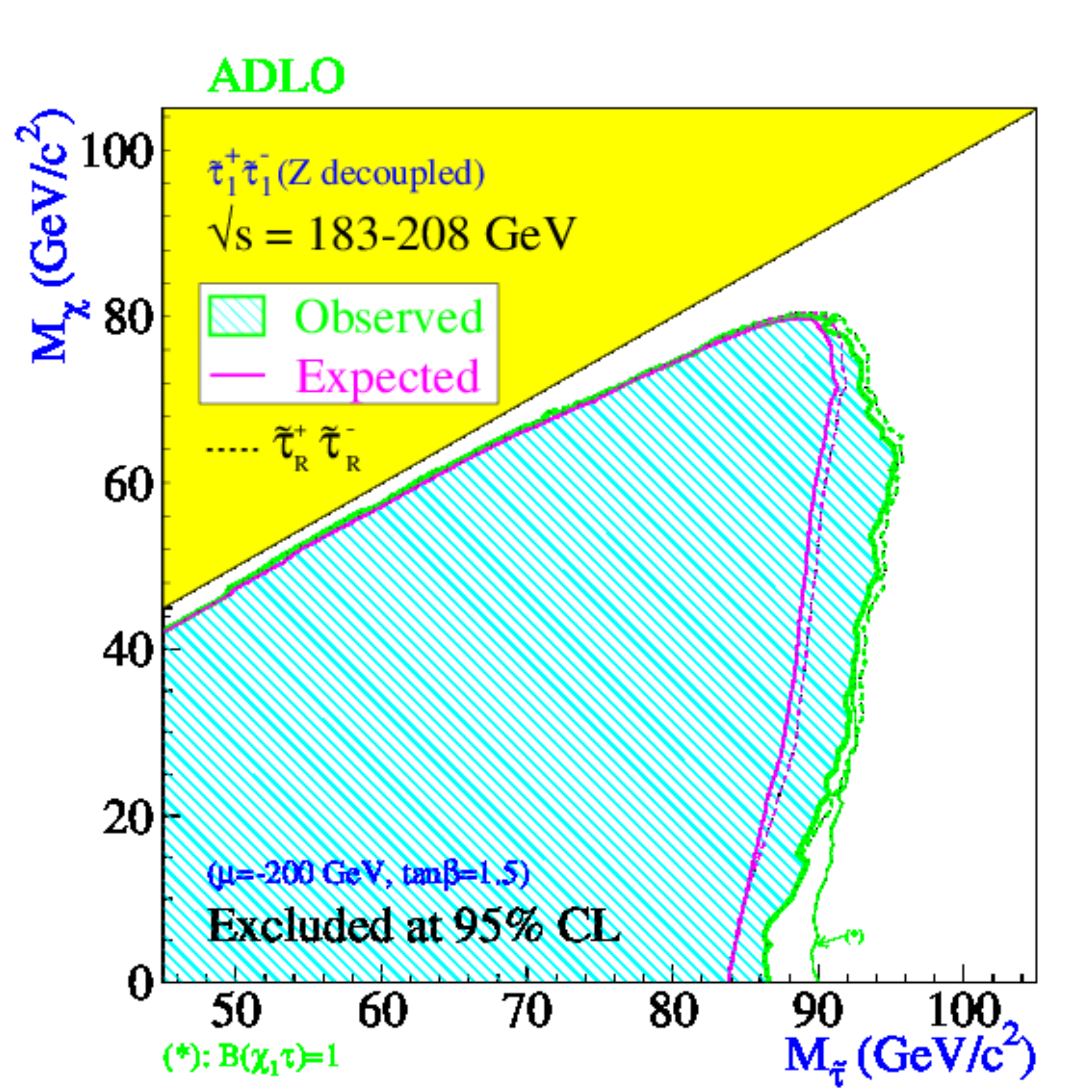}}
    \subcaptionbox{}{\includegraphics [bb=252 178 807 563,clip=true,scale=0.35]{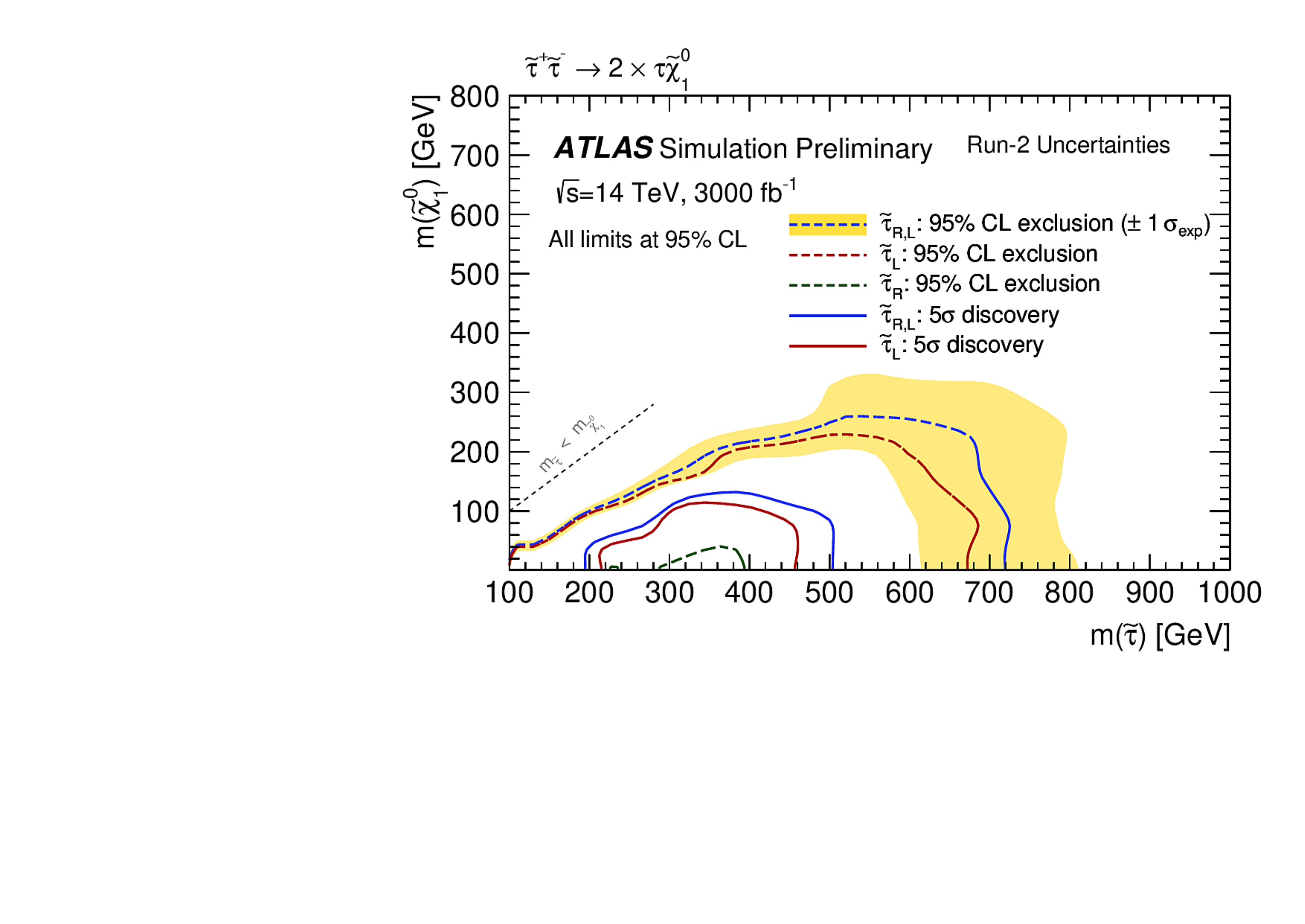}}
    \caption{(a): 95$\%$ CL exclusion limits for $\widetilde{\tau}$ pair production obtained combining data collected at the four LEP experiments with
      energies ranging from 183 GeV to 208 GeV. From~\cite{LEPSUSYWG/04-01.1}.
       (b): 95$\%$ CL exclusion and discovery potential for $\widetilde{\tau}$ pair production at the HL-LHC, assuming
    $\widetilde{\tau}_{L}^{+}\widetilde{\tau}_{L}^{-}$ + $\widetilde{\tau}_{R}^{+}\widetilde{\tau}_{R}^{-}$ production, $\widetilde{\tau}_{L}^{+}\widetilde{\tau}_{L}^{-}$ production or $\widetilde{\tau}_{R}^{+}\widetilde{\tau}_{R}^{-}$ production. From~\cite{ATLAS:2018diz}.}
    \label{lep_lhc_limits}
  \end{figure}

  The most model-independent limit on the $\widetilde{\tau}$ mass comes from the LEP experiments~\cite{LEPSUSYWG/04-01.1}.
  They set a minimum value that ranges from 87 to 93 GeV depending on the mass difference between the $\widetilde{\tau}$ and
  the neutralino, not smaller than 7 GeV. These limits, shown in figure~\ref{lep_lhc_limits}(a), are valid for any mixing 
  and any value of the model-parameters, other than the two masses explicitly shown in the plot.
  An analysis by the DELPHI experiment, targeted at low mass differences, excludes a $\widetilde{\tau}$ 
  with mass below 26.3 GeV, for any mixing, and any mass difference larger than the $\tau$ mass~\cite{Abdallah:2003xe}.

  At the LHC, ATLAS and CMS have determined limits on the $\widetilde{\tau}$ mass, analysing data from Run 1 and
  Run 2~\cite{ATLAS:2019gti,CMS:2019eln}.
  These limits, however, are only valid under certain assumptions. Both experiments assume $\widetilde{\tau}_{R}$
  and $\widetilde{\tau}_{L}$ to be mass-degenerate. This is a very unlikely scenario, the running of the $\widetilde{\tau}_{R}$
  and $\widetilde{\tau}_{L}$ masses from the GUT scale to the weak scale follows renormalisation 
  group equations with $\beta$-functions
  that are inevitably different for the two weak hyper-charge states.
  They also assume that there is no mixing between the weak hyper-charge eigenstates, which is again very improbable.
  The mixing will yield to cross section of the lightest physical state smaller than that of any unmixed state.
  Putting together $\widetilde{\tau}_{R}$ and $\widetilde{\tau}_{L}$ by adding the cross sections, ATLAS excludes
  $\widetilde{\tau}$ masses between approximately 120 and 390~\,GeV for a nearly mass-less neutralino\footnote{100$\%$
    decay to $\widetilde{\tau}$ and neutralino is assumed, as it is in the analysis presented in this paper}.
  Under the same conditions, CMS extends the lower limit to 90~\,GeV closing the gap with the LEP limit. Analysis of a pure
  $\widetilde{\tau}_{L}$ pair production set limits between 150 and 310~\,GeV from ATLAS data and up to 125~\,GeV from CMS; 
  both limits  again
  assume a nearly mass-less neutralino.

  The future HL-LHC should provide an improvement on the $\widetilde{\tau}$ limits provided by ATLAS and
  CMS, not only because of an increase of the luminosity but also because of an expected gain in sensitivity
  to direct $\widetilde{\tau}$ production due to the use of different analysis methods.
  Simulation studies have already been performed in both
  experiments~\cite{ATLAS:2018diz,CMS:2018imu}. Upper limits for $\widetilde{\tau}$ masses
  are indeed increased by about 300~\,GeV, but they suffer from the same constraints as the previous
  studies. ATLAS adds limits for pure $\widetilde{\tau}_{R}$ pair production, that could be considered
  the closest case to the physical lightest $\widetilde{\tau}$ since it is likely to be the lightest of the two
  weak hyper-charge states and is the one with the lower cross section. 
  These limits, presented in figure~\ref{lep_lhc_limits}(b),
  show that no discovery
  potential is expected in this case, only exclusion potential. 
  They do not have exclusion potential for $\widetilde{\tau}$ co-annihilation
  scenarios, a highly motivated scenario if SUSY is to provide a viable DM candidate:
  Such a scenario requires that the   $\widetilde{\tau}$-LSP mass difference is small, $\lesssim$ 10 GeV. 

  $\widetilde{\tau}$ searches at the ILC have been also performed in previous studies~\cite{Berggren:2013vna}.
  They assume an integrated luminosity of 500~\,fb$^{-1}$ at $\sqrt{s}=500$\,GeV and only used the $\mathcal{P}_{+-}$
  data sample.
  In the current study,
  both   $\mathcal{P}_{+-}$ and  $\mathcal{P}_{-+}$ 
  data are used,
  and the luminosity is increased
  to the one corresponding the the foreseen running scenario, 1.6~\,ab$^{-1}$ at each of the beam-polarisation
  combinations.
  The previous study was aimed at scanning the entire $M_{LSP} - M_{NLSP}$ plane, and doing so for several different
  NLSP candidates.
  Quite generic cuts were therefore used, and were optimised ``on-the-fly'' at different points. 
  More specifically, the limits presented in that study do not have a dedicated
  analysis for low mass differences between the $\widetilde{\tau}$ and the LSP, $\Delta M$, and are only
  valid down to $\Delta M$ 3-4~\,GeV. The exclusion limit goes up to 240\,GeV with a discovery potential
  up to 230\,GeV for large mass differences.

  A recent study of slepton production at ILC and,  in general, future e+e- colliders 
         can be found in \cite{Baum:2020gjj}.  They have demonstrated
         that these colliders would be able to probe most of the kinematically
         accessible parameter space with only a few days of data, with the
         capability of discovering/excluding new physics that would evade
         detection at the LHC.
  

\section{Conditions and tools\label{sec:tools}}
  The study was done assuming an integrated luminosity of 1.6~\,ab$^{-1}$ at $\sqrt{s}=500$\,GeV
  for each of the beam polarisations   $\mathcal{P}_{+-}$ and  $\mathcal{P}_{-+}$,
  according to the H-20
  running scenario for the ILC500 \cite{Barklow:2015tja}\footnote{The H20 scenario is defined as: $\sqrt{s}$=500\,GeV, total integrated
  luminosity 4\,ab$^{-1}$ with 1.6\,ab$^{-1}$ for $\mathcal{P}_{+-}$ and $\mathcal{P}_{-+}$ 
  , 0.4\,ab$^{-1}$ for  $\mathcal{P}_{--}$ and $\mathcal{P}_{++}$.}.
  The study assumes R-parity conservation and a 100$\%$ decay of the $\widetilde{\tau}$ to ${\tau}$
  and the lightest neutralino ($\chi^0_1$), the LSP in this case.
    In order to select the worst scenario, the $\widetilde{\tau}$ mixing angle was set to 53 degrees.
  This mixing angle corresponds to the lowest cross section in the case of un-polarised beams, due to 
  the suppression of the s-channel with $Z$ exchange
  in the $\widetilde{\tau}$ pair production.
  In section \ref{sec:worst}, we will show that this mixing angle in fact also corresponds to the worst case
  at ILC operating according to the H20 scenario.
  The generated background event samples were those of the 
  standard ``IDR'' production~\cite{ILDConceptGroup:2020sfq}. They were generated
  with {\tt Whizard} 1.95~\cite{Kilian:2007gr}, and contain {\it all} standard model processes with up to six fermions in
  the final state. 
  Beam-spectra and the amount of photons in the beams were simulated with {\tt GuineaPig} \cite{Schulte:1999tx}.
  Detector simulation and reconstruction were done on the Grid using {\tt DDSim} \cite{Petric:2017psf} 
  and {\tt Marlin} \cite{Gaede:2006pj}.
  The Grid production was done by the ILD production team using the {\tt Dirac} \cite{Tsaregorodtsev:2008zz} system.
  The {\tt SGV} fast detector simulation~\cite{Berggren:2012ar}, adapted to the ILD  concept~\cite{ILDConceptGroup:2020sfq} at ILC, 
  was used for detector simulation and event
  reconstruction for signal events. 
  These events were generated
  with {\tt Whizard} v2.8.5, interfaced to {\tt Tauola} \cite{Jadach:1990mz}.
  {\tt Tauola} simulates the ${\tau}$-decays taking into account the ${\tau}$ polarisation in the products.
  The same beam-spectrum as was used for the fully simulated background samples was also used for the signal sample.
  Both the signal and background samples were analysed using the tools
  included in {\tt SGV}, and the
  relevant information of the reconstructed events were written to Root files.

  At ILC spurious events are expected to be present in each beam-crossing. At the ILC
  with $\sqrt{s}=500$\,GeV an average of 1.5 low $\it{P_{T}}$ hadron events 
  from $\gamma\gamma$ interactions 
  is expected per bunch crossing. 
  A number ($\mathcal{O}(10)$ ) of  electron-positron pairs from beam-beam interactions is also expected to reach the
  tracking system of the detector in each bunch crossing.
  The $\gamma\gamma$ interactions were generated either with {\tt Pythia} 6.422~\cite{Sjostrand:2006za} (if  $M_{\gamma\gamma}$ > 2 GeV)
  or a dedicated generator \cite{Chen:1993dba} for $\gamma\gamma$ interactions (otherwise).
  The electron-positron pairs from beam-beam interactions were generated with
  {\tt GuineaPig}. A pool of such events were created, and random events picked from
  the pool were added to each physics event during full simulation.
  The {\tt SGV} simulation did not implement this mechanism, instead reconstructed
  objects coming from overlay were extracted from random background events and
  overlaid on the signal events at ntuple level.

  Final significance values were computed adding the contribution of
  both polarisations.
  The signal production cross-section depends on the beam polarisation.
  Even more so, the background levels are expected
  to be quite different for the two cases,
  mainly because the strongly polarisation dependent $e^+e^- \rightarrow W^+W^-$ process is one of the most
  important sources of background.
  Because of this the samples are 
  weighted by the likelihood ratio (LR) test statistic \cite{Read:2000ru,Neyman:1933wgr}.
  As the experiment is a pure event-counting one, the LR statistic has the simple form
  \begin{equation}
    N_{\sigma} = \frac{\sum^{n_{samp}}_{i=1} s_i \ln{(1+s_i/b_i)}}{
      \sqrt{ \sum^{n_{samp}}_{i=1} n_i \left [\ln{(1+s_i/b_i)} \right]^2}}
    \label{eq:LR}
  \end{equation}
  where $s_i$ and $b_i$ are expected number of signal and background events in each sample,
  and $N_{\sigma}$ is the significance of the test, expressed as the equivalent number of standard deviations in
  the Gaussian limit. Depending on which hypothesis is tested (exclusion or discovery), $n_i$ is either $s_i +b_i$ (exclusion),
  or $b_i$ (discovery).

 \begin{figure}[htbp]
    \centering
    \subcaptionbox{}{\includegraphics [scale=1.0]{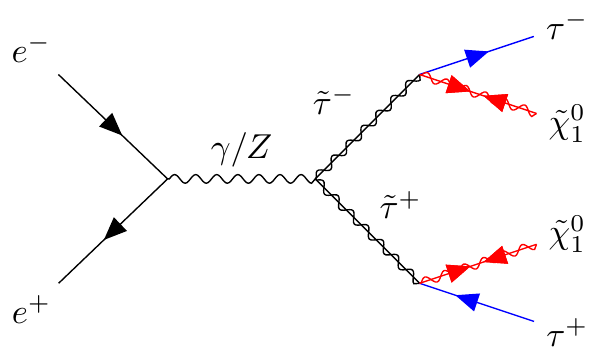}}

    \subcaptionbox{}{\includegraphics [scale=1.0]{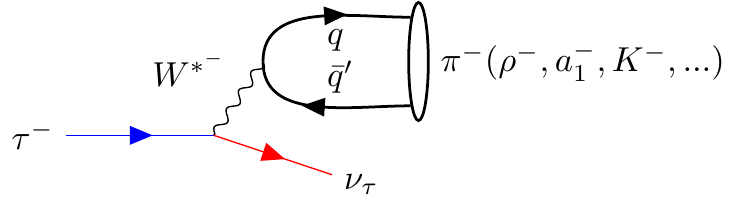}}
    \subcaptionbox{}{\includegraphics [scale=1.0]{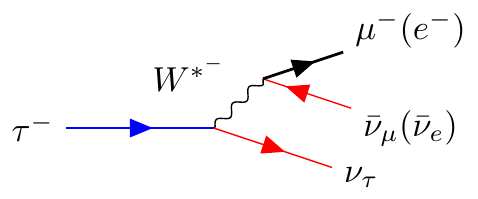}}
    \caption{(a) $\widetilde{\tau}$ pair production and decay to $\tau$ and $\widetilde{\chi}^0_1$; (b) hadronic $\tau$ decay; (c) leptonic $\tau$ decay.
             Blue lines indicates the $\tau$, red lines indicate undetectable final state particles, while thick black lines indicate the
             detectable final state particles.}
    \label{fig:signdiagrams}
  \end{figure}
  Well identifying $\tau$:s is obviously a key requirement for this analysis.
  To do this, we use the DELPHI $\tau$ finder ~\cite{Abdallah:2003xe}.
  This algorithm was particularly developed to identify $\tau$:s in low multiplicity
  events. 
  It iterative builds $\tau$ candidates from all possible combinations of charged tracks, starting with single tracks,
  then in each iteration adding tracks to existing combinations, always retaining the combination yielding the
  lowest mass, and requiring that the mass is below 2 GeV. Characteristics of $\tau$ decays are also taken into account,
  so that e.g. an identified muon is not allowed to be grouped with another track. The charged track grouping is terminated
  when no more groups with mass below 2 GeV can be made. Neutrals are then added to the found candidates, as long as the mass
  remains below 2 GeV - any neutrals that can not be grouped are left as belonging to the ``rest-of-event''
  class. To allow for tracks not being found, it is {\it not}, at this stage, required that the groups have charge $\pm$ 1.

\section{Signal and background \label{sec:signandbck}}
\subsection{Signal characterisation \label{sec:signcharac}}
  Assuming R-parity conservation and assuming that the $\widetilde{\tau}$ is the NLSP, $\widetilde{\tau}$'s
  will be produced in pairs via $Z^{0}$/$\gamma$ exchange in the s-channel and they will decay to
  a ${\tau}$ and an LSP (assuming mass differences above the mass of the ${\tau}$, as is done
    in this study).
  The LSP, as already mentioned, is stable and weakly interacting, hence it will leave the detector
  without being detected. The ${\tau}$, with a proper lifetime of 2.9 x 10$^{-13}$ s, will decay
  before leaving any signal in the detectors. The only detectable activity in the signal events is therefore
  the decay products of the two ${\tau}$'s.
  Figure~\ref{fig:signdiagrams}(a) shows the diagram of the $\widetilde{\tau}$ production and decay,
  Figures~\ref{fig:signdiagrams}(b) and (c) show the diagrams of the subsequent $\tau$ decays, in the hadronic
  and leptonic modes, respectively.
  Signal events are therefore characterised by a large missing energy and momentum, not only due to the
  invisible LSPs but also to the neutrinos from both ${\tau}$-decays. Since the $\widetilde{\tau}$'s are scalars
  and hence isotropically produced, these events have a large fraction of the detected activity in the
  central region of the detector. 
  \begin{figure}[htbp]
    \centering
    \subcaptionbox{}{\includegraphics [scale=1.0]{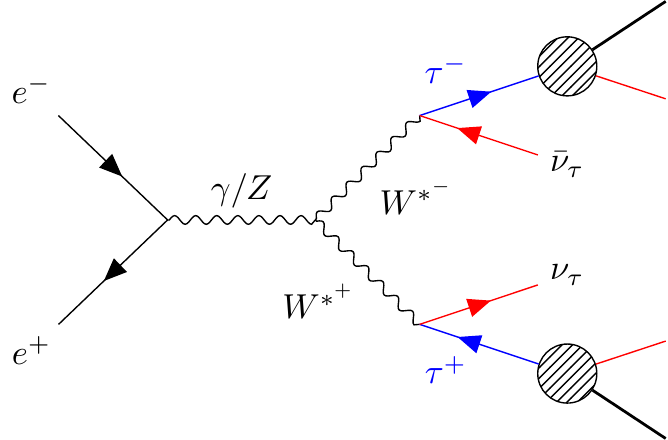}}
    \subcaptionbox{}{\includegraphics [scale=1.0]{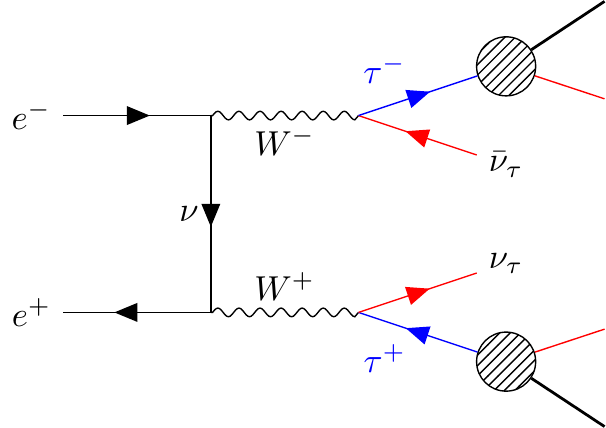}}

    \subcaptionbox{}{\includegraphics [scale=1.0]{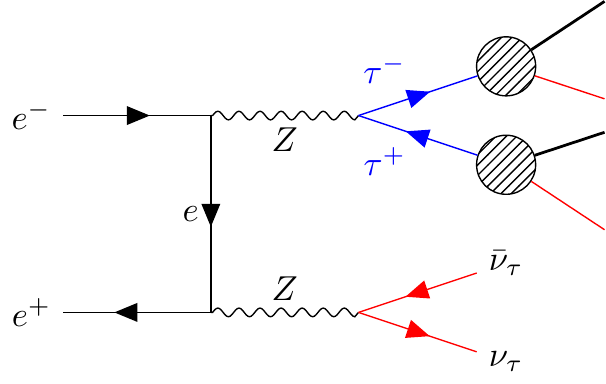}}
    \caption{Main irreducible background sources: 
(a) s-channel $WW$ production with both $W$:s decaying to $\tau$ and $\nu$; 
(b) t-channel $WW$ production with both $W$:s decaying to $\tau$ and $\nu$; 
(c) t-channel $ZZ$ production with one $Z$ decaying to $\tau$:s, the other to $\nu$:s. 
The same colour-coding is as in Figure \ref{fig:signdiagrams} is used, and the various $\tau$ decays modes (cf. Figures~\ref{fig:signdiagrams}(b) and (c))
are collectively represented by the hatched circles.
    \label{fig:irredbckdiagrams}}
  \end{figure}
  The $\widetilde{\tau}$'s must also be rather heavy, so they will not have a large boost in the lab-frame,
  and since the LSP is also quite heavy, the direction of the $\widetilde{\tau}$ does not  strongly correlate to that of
  the visible $\tau$ after the  decay. 
  As a consequence the two $\tau$-leptons are expected to go in directions quite independent
  of each other resulting in events with un-balanced transverse momentum, large angles between the 
  two $\tau$-lepton directions and zero forward-backward asymmetry.
  These properties are  however not necessarily present in any event - the two $\tau$'s could accidentally happen
  to be back to back, for example.


\subsection{Main background sources\label{sec:mainbck}}
  The main sources of background, given the generic signal topology, i.e. two $\tau$'s and an unseen
  recoil system, are SM processes with real or fake missing energy. They can be classified into
  ``irreducible'' and ``almost irreducible'' sources. The first are events with two $\tau$'s
  and real missing energy, i.e. neutrinos. The main contribution to this group are $ZZ$ events
  with one $Z$ decaying to two neutrinos and the other to two $\tau$'s, and fully leptonic $WW$ events,
  where both the $W$'s decays to $\tau$ and neutrino.
  The dominating diagrams are shown in Fig. \ref{fig:irredbckdiagrams}.
  In principle, $ZWW$ and $ZZZ$ events decaying to two $\tau$'s and four neutrinos also contribute, but
  are of less concern due to their low cross sections.
  Nevertheless, these process are included in the simulated background samples.
\begin{figure}[htbp]
    \centering
    \subcaptionbox{}{\includegraphics [scale=1.0]{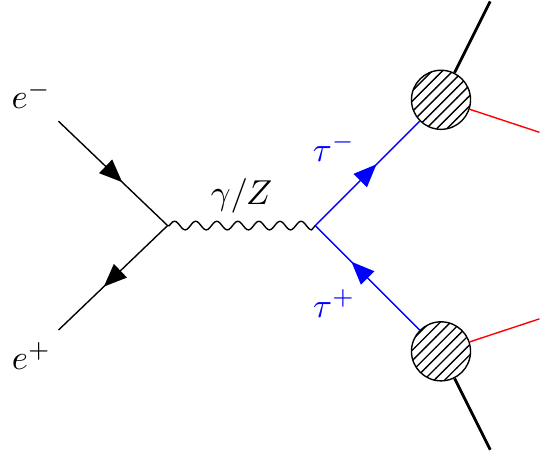}}
    \subcaptionbox{}{\includegraphics [scale=1.0]{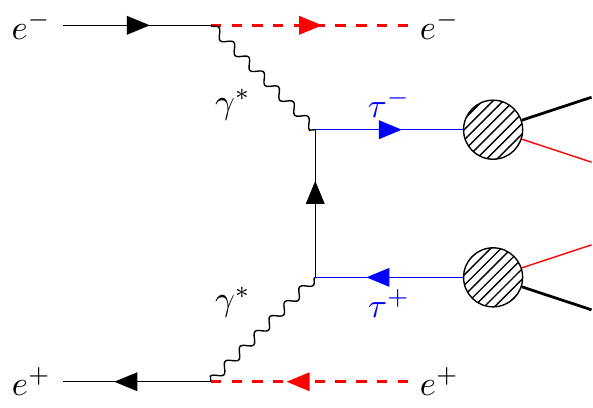}}
    \caption{Main background sources: (a) $\tau$ pair production and decay; 
             (b)  $\tau$ pair production in $\gamma\gamma$ events.
             In (b), the thick dashed red lines indicate that while the outgoing
             electrons or positrons are in principle detectable, this diagram contributes to the background only in the case that they are
             deflected so little that they escape detection by remaining inside the beam-pipe, and hence contribute to (fake) missing energy.}
    \label{fig:redbckdiagrams}
  \end{figure}

  The second group of events are those which are not really two $\tau$'s and neutrinos, but after reconstruction look very similar.
  They are events with two soft $\tau$-jets, with two other leptons plus true
  missing energy or two $\tau$'s plus fake missing energy.
  The main sources for events with true missing energy in this group are on one hand $\tau$
  pair production, with the $\tau$'s decaying such that most energy goes to the neutrinos, (Figure \ref{fig:redbckdiagrams}(a)), 
  and on the other hand, $ZZ$ or $WW$ decaying to two neutrinos and at least one lepton other than $\tau$'s, i.e. the
  type of processes shown in Figure \ref{fig:irredbckdiagrams}, but with one, or both, $\tau$'s replaced by muons or electrons.
  The background with fake missing energy comes mainly from $\tau$ pair
  production with Initial State Radiation (ISR) at very low angles, events with two $\tau$'s and
  two very low angle electrons (below the acceptance of the detector) in the final state (Figure \ref{fig:redbckdiagrams}(b)) and events 
  where two $\tau$'s are produced
  by a $\gamma\gamma$ interaction and not from an $e^+e^-$ one  ; in the latter case there
  is not really missing energy but the assumption that the initial energy is
  the energy of the incoming electron and positron is wrong.


\section{Analysis\label{sec:analysis}}
\subsection{General cuts\label{sec:gencuts}}
  
  Taking into account the signal signature and the main background sources,
  different cuts have been designed in order to separate the signal from
  the background.
    \begin{figure}[htbp]
    \centering
      \subcaptionbox{}{ \includegraphics [width=0.45\textwidth]{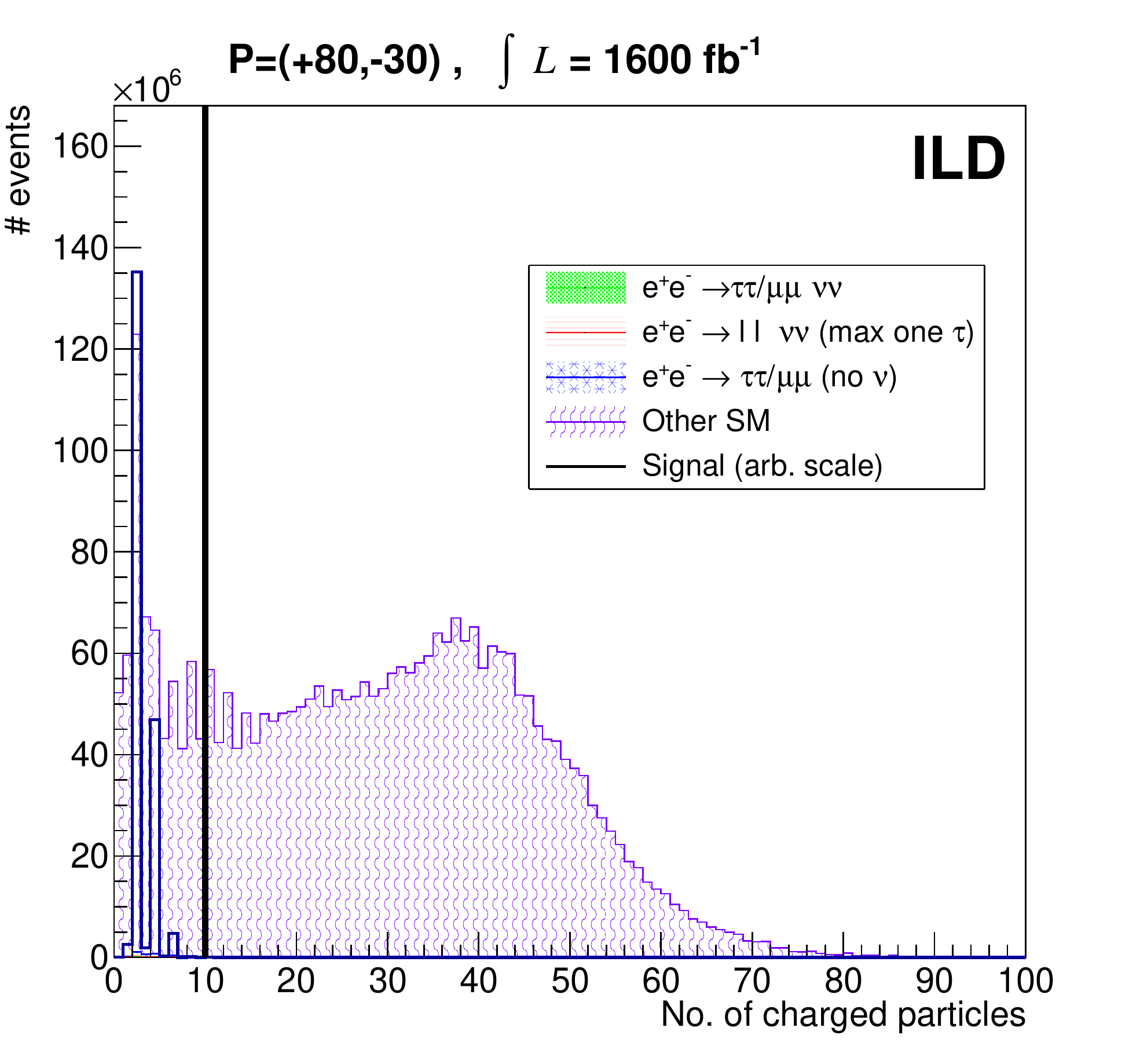}}
   \subcaptionbox{}{\includegraphics [width=0.45\textwidth]{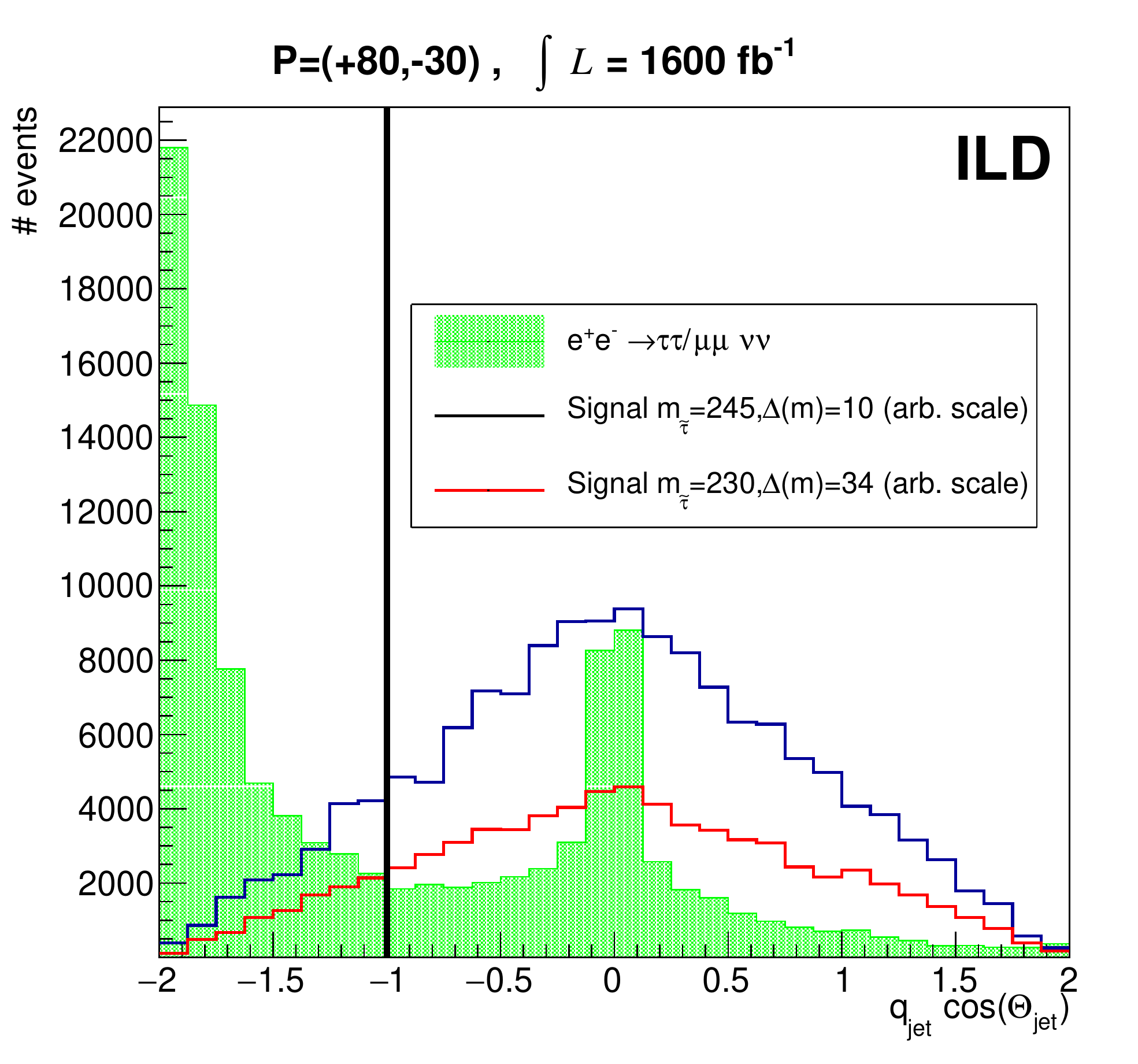}}
   \caption{Distributions for background and signal (the latter on an arbitrary scale).
     (a): the number of charged particles. For clarity, the $\gamma\gamma$ background has been suppressed
     in this figure.
     (b) : $q_{jet} \cos{\theta_{jet}}$ after the cut on  the number of charged particles. Only the irreducible $WW \rightarrow l \nu l \nu$
     is shown, since this background is what this cut is designed to reduce. Events above the vertical lines are accepted.}
    \label{fig:nbch_qcos_dm34}
  \end{figure}
  The study was focused on small differences between the $\widetilde{\tau}$ and LSP
  masses\footnote{Larger mass differences were also analysed in order
  to cross-check and try to improve the limits from the previous studies.},
  $M_{\tau}$ $<$ $\Delta M$ $<$ 11\,GeV. This class of signal events are quite similar to
  high cross-section multi-peripheral $\gamma\gamma$ events, with one important difference, namely that 
  in the  $\gamma\gamma$ events the final state also includes the beam-remnant electrons and positrons.
  These are scattered by some low angle, so that demanding absence of signal in the calorimeter closest to the beam pipe
  (the BeamCal) strongly reduces this source of background, and was required as a pre-selection 
  step before applying the following cuts. 
  To reject tracks from low invariant mass  $\gamma\gamma$ events and from beam-beam interactions (``overlay tracks'') 
  occurring in the same bunch-crossing as the physics event, a selection of tracks is performed, details of
  which are given in Sect. \ref{sec:overlcuts}. 

  All quantities mentioned in the following are to be understood as being calculated using reconstructed
  objects passing these cuts. Tables \ref{tab:cutsflow_dm34} and \ref{tab:cutsflow_dm10} show the cut-flow for
  the cuts applied at two model-points in the order the cuts are applied in the discussion that follows.
  All figures in this section will show background and signal distributions for various quantities that
  cuts are applied to at the point where the cut on the shown quantity will be the next one to be done,
  i.e. for events passing all previous cuts.

\begin{figure}[htbp]
    \centering
      \subcaptionbox{}{ \includegraphics [width=0.45\textwidth]{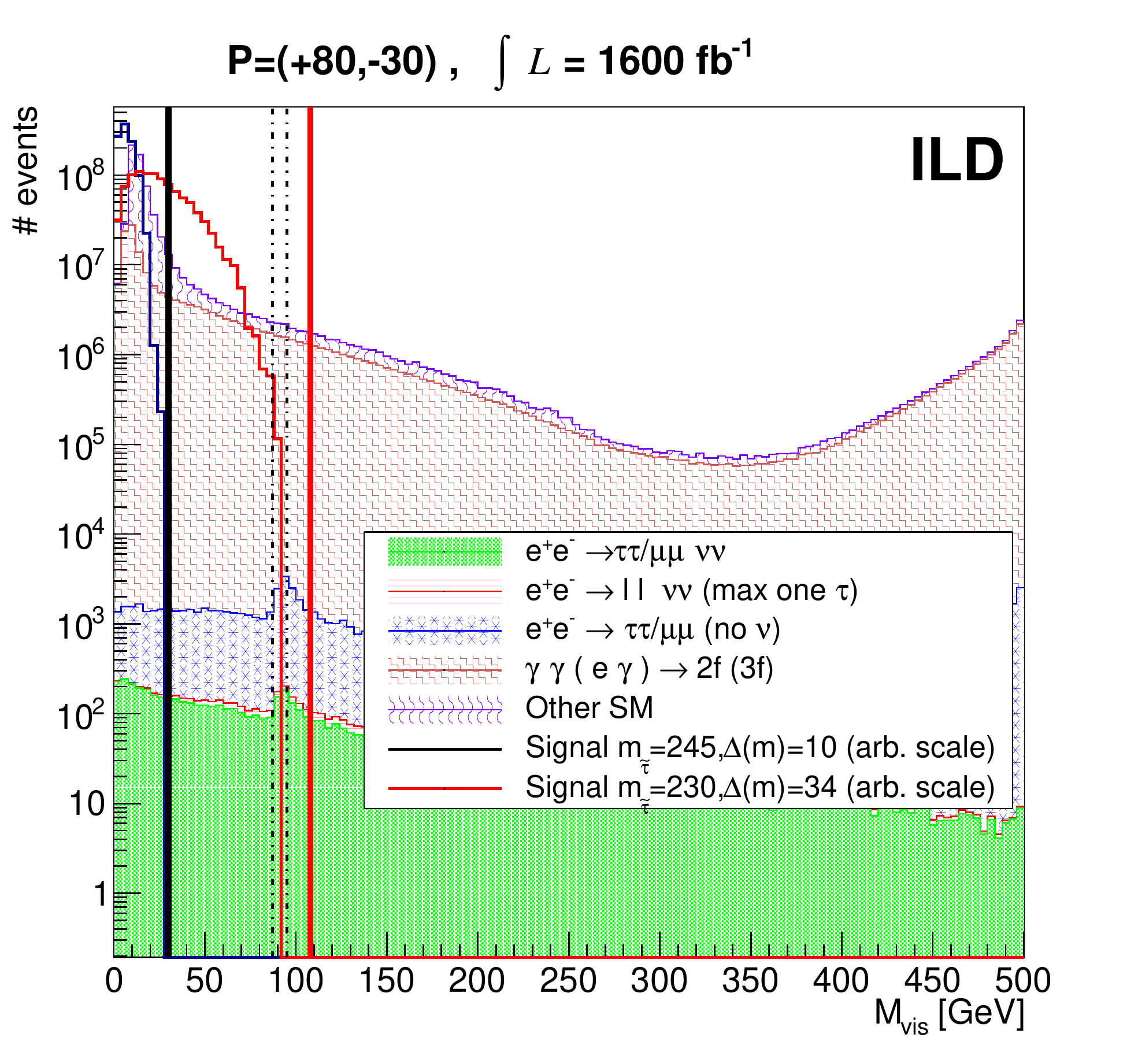}}
   \subcaptionbox{}{\includegraphics [width=0.45\textwidth]{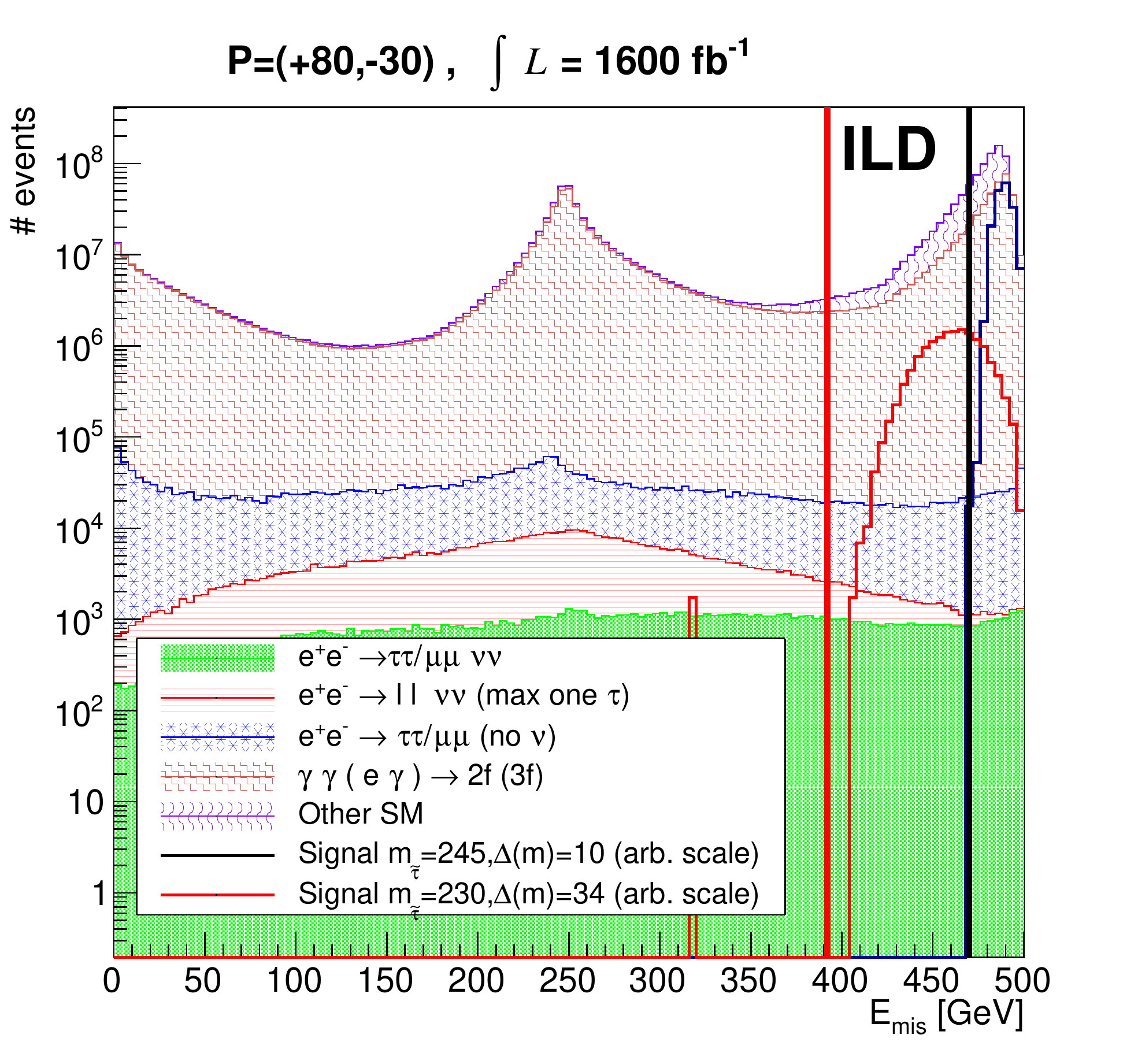}}
   \caption{Distributions of the visible mass (a) and the missing energy (b), for two signals (arbitrary scale),
     and background, after previous cuts have been applied (cf. Tabs. \ref{tab:cutsflow_dm34} and  \ref{tab:cutsflow_dm10}).
     Events below (above) the vertical
     lines are accepted in a(b) at the model points indicated by the colour of the lines.
     In the $M_{vis}$ distribution, also events between the dashed lines around $M_Z$ were rejected.}
    \label{fig:emis_zpeak_dm34}
  \end{figure}

  Initially, cuts are applied on properties that the signal often {\it does not} have,
  while some backgrounds {\it do have}:
  The multiplicity of the event can
  be constrained taking into account that the visible part of the $\widetilde{\tau}$ decays comes only from
  the decays of the two $\tau$'s and maybe an ISR photon. 
  For that reason
  the number of charged particles in the event is required to be between 2 and 10.
  This cut
  removes practically all hadronic background, see Fig. \ref{fig:nbch_qcos_dm34}(a).
$WW$ events with each of the $W$'s decaying to a lepton and a neutrino are
  highly forward-backward asymmetric; they can be effectively removed by requiring the sum of the product of the
  charge and the cosine of the polar angle, $ q \cos{ \theta_{jet}}$, of the two most energetic jets to be above -1.0,
  see Fig. \ref{fig:nbch_qcos_dm34}(b).
  $ZZ$ events with one $Z$ decaying to two neutrinos and the second one to a electron or muon
  pair are highly suppressed demanding $|M_{vis} - M_Z| > 4$ GeV, see Fig. \ref{fig:emis_zpeak_dm34} (a).

  
  The cuts mentioned in the previous paragraph does not depend on
  the properties of the signal model point. However, all following cuts applied will depend on the model-point considered,
  and so will depend on either, or both, of $M_{\widetilde{\tau}}$ and $M_{LSP}$.

\begin{figure}[htbp]
    \centering
\includegraphics [scale=1.0]{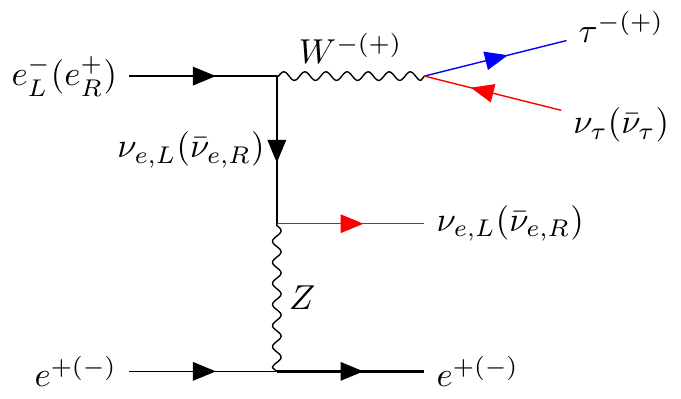}
    \caption{
     Single $W$ production with  
              the $W$ decaying to $\tau$ and $\nu$.
      The labels indicate the only beam-polarisations for which this process is possible:
             either left-handed electrons or right-handed positrons. 
    \label{fig:singleWdiagram}}
  \end{figure}

  The first group of cuts contain those in properties that the
  $\widetilde{\tau}$-events {\it must} have. Since the two LSPs from the
  $\widetilde{\tau}$-decays are invisible to the detector, signal events
  have to have a missing energy, $E_{miss}$, greater than $2 \times M_{LSP}$.
     \begin{figure}[htp]
    \centering
    \subcaptionbox{}{\includegraphics [width=0.45\textwidth]{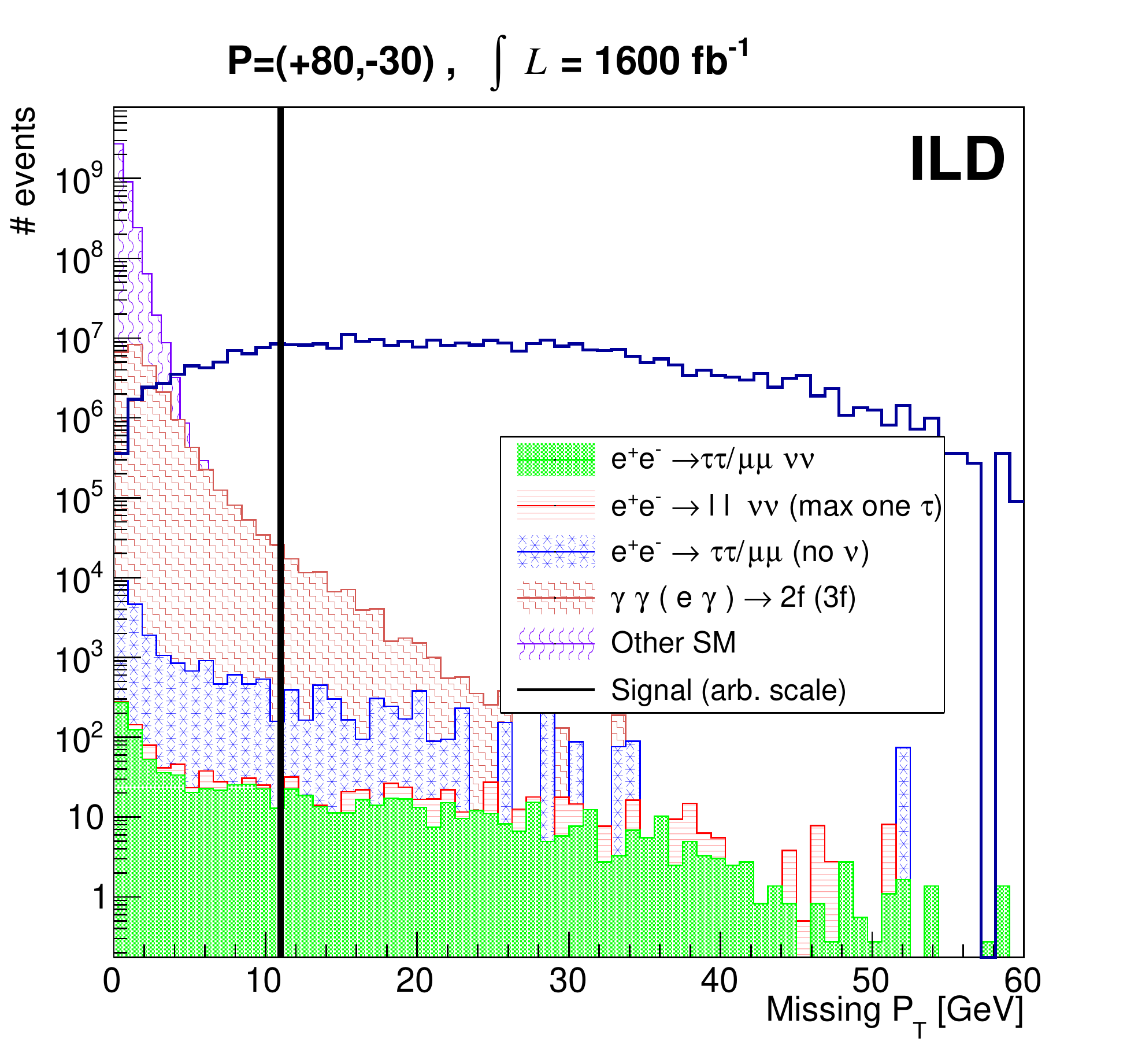}}
     \subcaptionbox{}{ \includegraphics [width=0.45\textwidth]{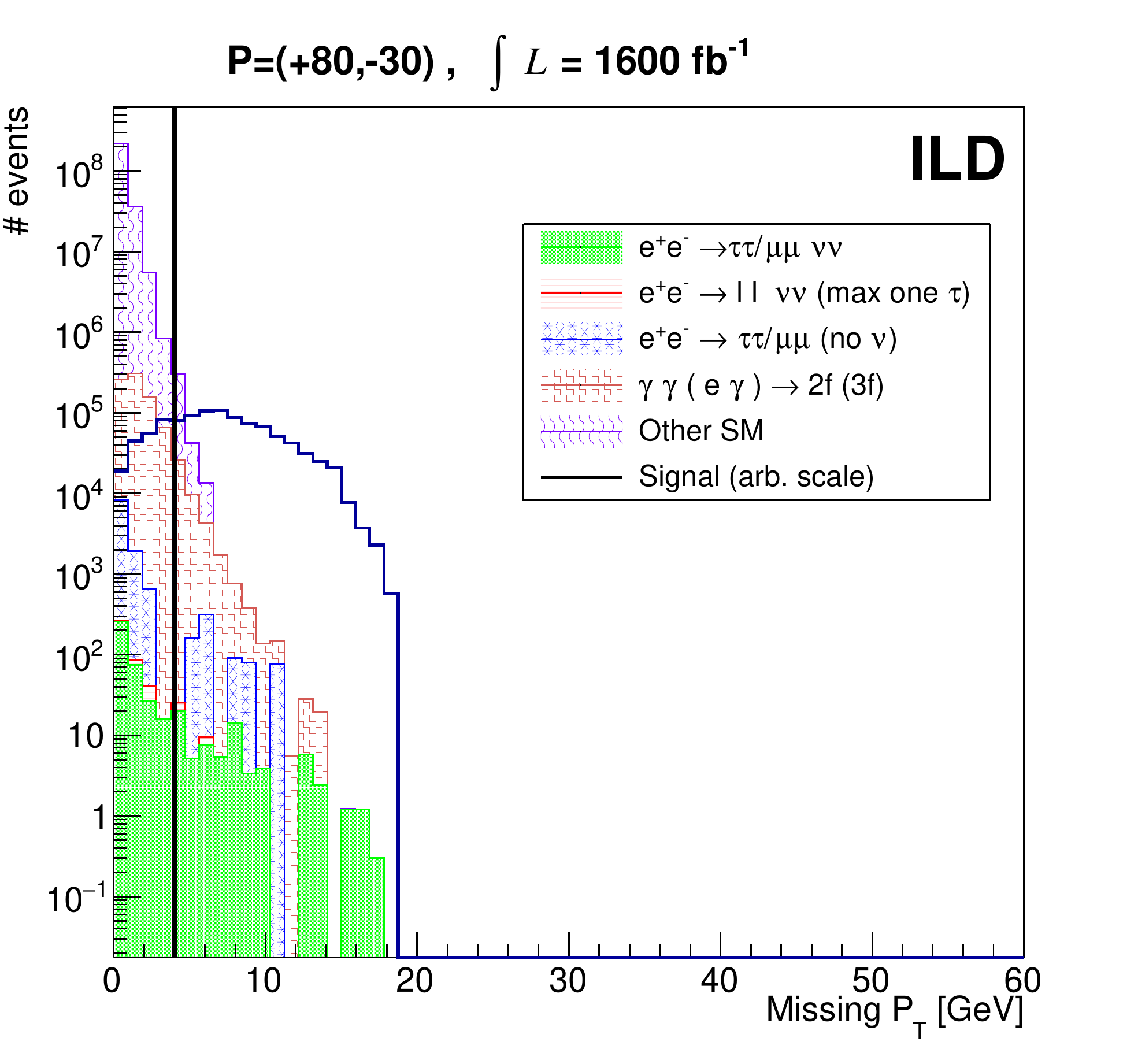}}
    \caption{
     Distributions of the missing transverse momentum for  (a) M$_{\widetilde{\tau}}$ = 230 GeV $\Delta(M)$ = 34 GeV,
     and (b)  M$_{\widetilde{\tau}}$ = 245 GeV $\Delta(M)$ = 10 GeV. The two signals are on arbitrary scale,
     and all previous cuts have been applied (cf. Tabs. \ref{tab:cutsflow_dm34} and  \ref{tab:cutsflow_dm10}).
     Events above the vertical
     lines are accepted.}
    \label{fig:ptmiss_dm34_dm10}
  \end{figure}
  Likewise, the visible mass, $M_{vis}$ can not be greater than $E_{CMS} - 2 \times M_{LSP}$.
  Therefore, events should fulfil $M_{vis} < E_{CMS} - 2 \times M_{LSP}$ (Fig. \ref{fig:emis_zpeak_dm34} (a)) and
  $E_{miss} > 2 \times M_{LSP}$ (Fig. \ref{fig:emis_zpeak_dm34} (b))
  to be considered further.
  Also a
  cut in the maximum total momentum, smaller than 70$\%$ the beam momentum,
  is applied for the same reason: $P_{tot} < 0.70 E_{beam}$. 
  There should only be 
  2 or 3 clusters found by the Jade algorithm (with $y_{cut}$=0.02)
  and a total seen event charge should be between
  -1 and 1. 
  A specific algorithm for $\tau$-identification was also applied.
  This algorithm consists in a first set of conditions requiring to have a pattern
  of charged tracks typical for $\tau$-decay, {\it viz.} exactly two $\tau$-jets
  (obtained with the DELPHI tau-finder) with
  1 or 3 charged particles in each charged jet, 
  jet-charge $\pm$1, and opposite charge between both jets. 
  These two jets could be accompanied by further particles,
  that are not compatible with the requirements imposed by the DELPHI tau-finder
  to be considered as $\tau$-jets.
  A further set of conditions on the jets
  is devoted to the reduction of background from sources with
  leptons not from $\tau$-decays.
  The background of single $W$,
  with the $W$ decaying to $\tau$ and neutrino, see Figure \ref{fig:singleWdiagram},
  can be reduced by a cut depending on beam-polarisation:
  this process can only occur if at least one of the beams has the
  correct polarisation.
  Since the degree of polarisation of the positron beam is lower than that of the electron beam,
  this background is more likely to yield an electron as the beam-remnant for $\mathcal{P}_{+-}$, a
  positron for $\mathcal{P}_{-+}$. We therefore remove events with an electron(positron) $\tau$ candidate in the
   $\mathcal{P}_{+-}$ ( $\mathcal{P}_{-+}$) samples.
  This background
  together with the background from $WW \rightarrow e\nu_{e} \mu\nu_{\mu}$ and
  from $\gamma\gamma$ events with a beam-remnant deflected to larger
  angles is further reduced by rejecting those events in which the most energetic
  jet consists of a single electron. 
  The two charged jets were also required to
  not be made by single leptons with the same flavour.
  These selections reduce the signal efficiency to 38$\%$ but with
  a reduction of the background of the order of 94$\%$, depending on the
  region of the SUSY parameter space.
  
  Since
  the $\widetilde{\tau}$-decay is a two body decay, it is possible to
  determine the maximum and minimum momentum of each of the  decay products as
  a function of the $\widetilde{\tau}$ mass, the mass of LSP
  and the centre-of-mass energy of the collider. A cut in the minimum momentum
  can not be applied due to the presence of neutrinos in the $\tau$ decay, with
  the corresponding decrease of observable momentum. The maximum value can be used even if
  it is smeared by the missed neutrinos. The expression for the maximum
  jet momentum is given by:

  \begin{equation}
    P_{max} = \frac{\sqrt{s}}{4}\left(1-\left ( \frac{M_{LSP}}{M_{\widetilde{\tau}}} \right )^2\right)\left(1+\sqrt{1-\frac{4M_{\widetilde{\tau}}~^2}{s}}\right)
    \label{eq:pmax}
  \end{equation}
  and events with the higher jet momentum above this limit are excluded from the further analysis at the
  given model point.
  Excluding the cut applied by the $\tau$-identification algorithm, the signal
  efficiency for each of the cuts is at least 95$\%$ at all model points.
  
    \begin{figure}[htbp]
    \centering
    \subcaptionbox{}{\includegraphics [width=0.45\textwidth]{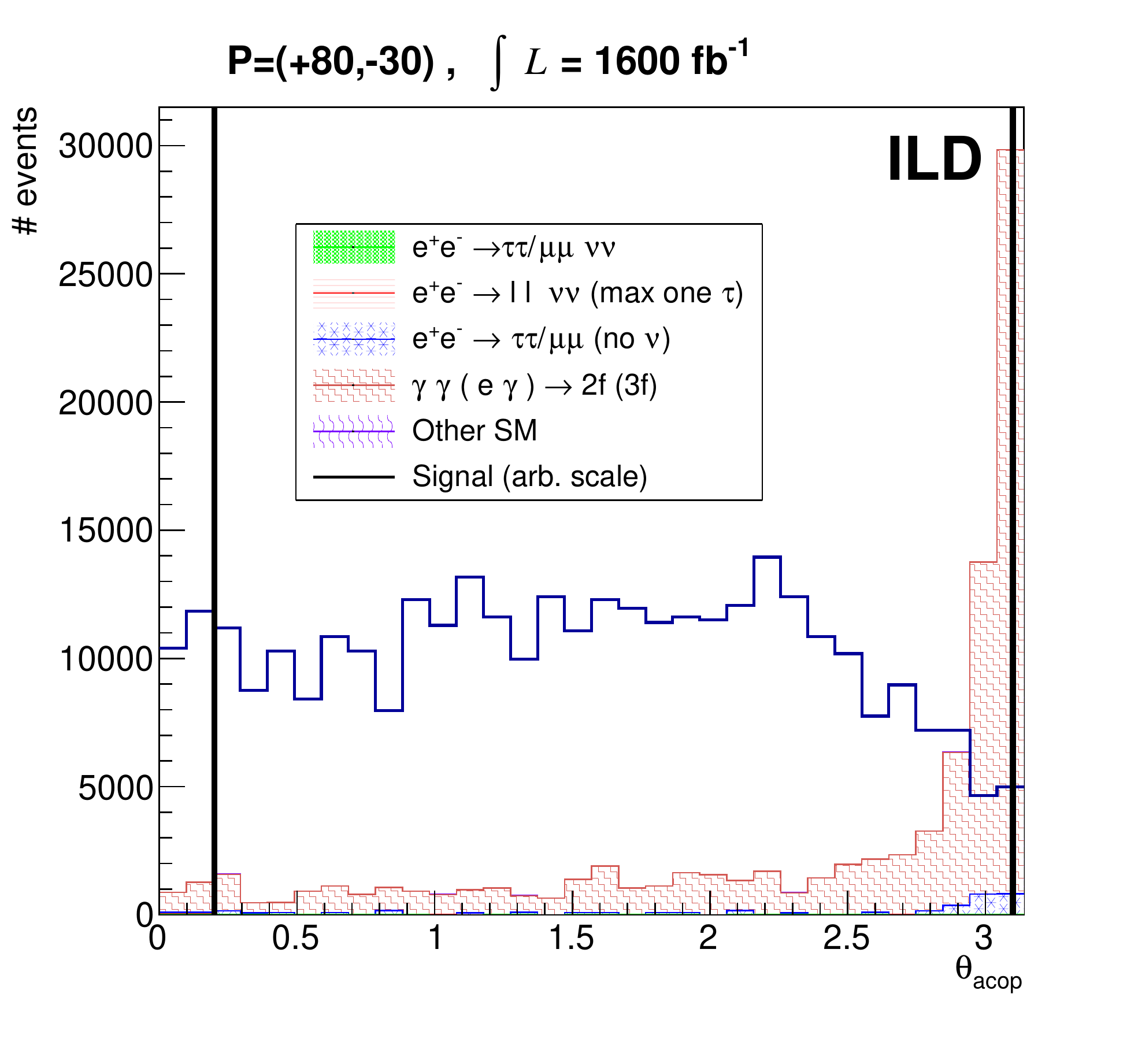}}
    \subcaptionbox{}{ \includegraphics [width=0.45\textwidth]{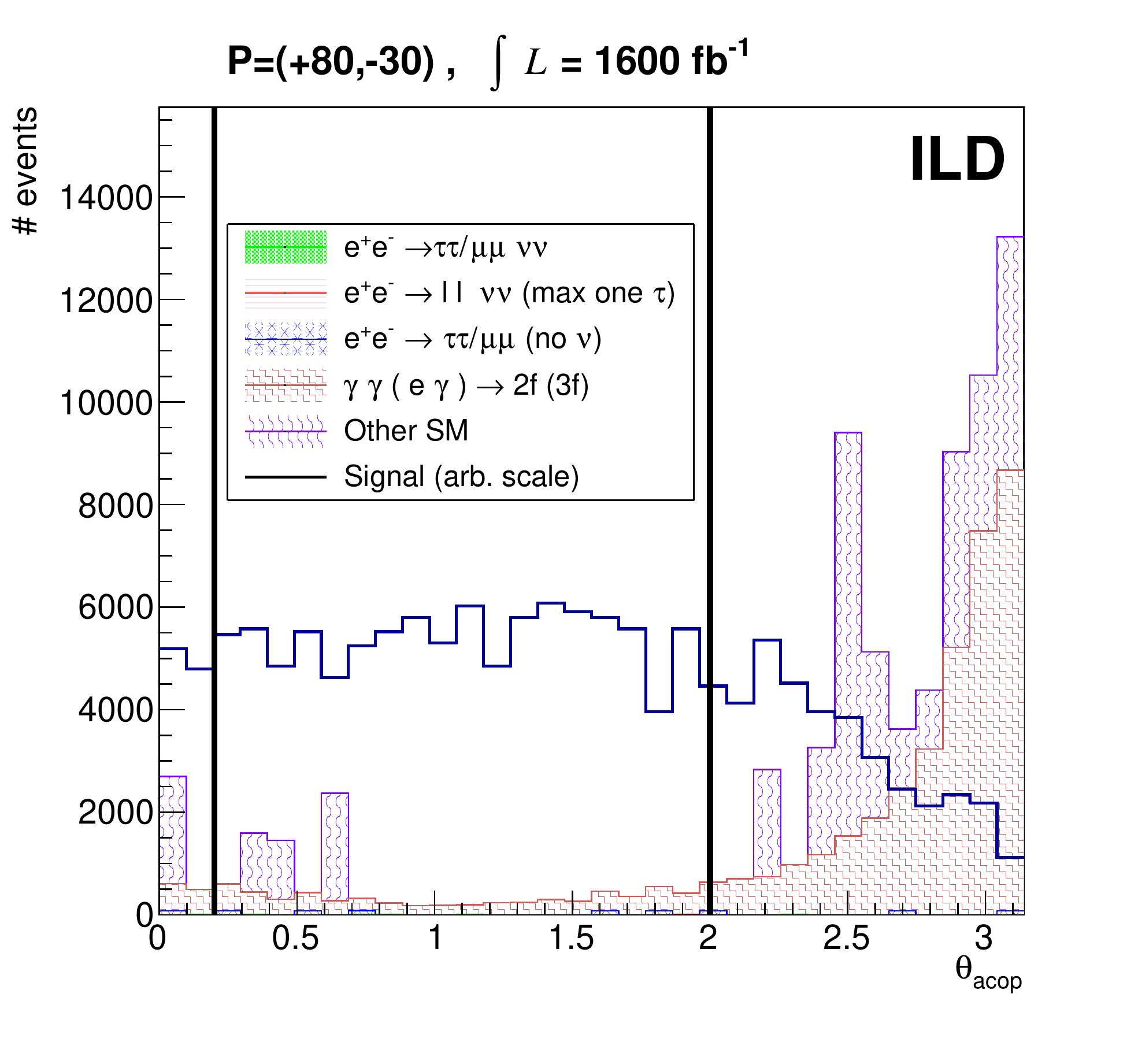}}
   \caption{
     Distributions of the acoplanarity angle between the two jets for  (a) M$_{\widetilde{\tau}}$ = 230 GeV $\Delta(M)$ = 34 GeV,
     and (b)  M$_{\widetilde{\tau}}$ = 245 GeV $\Delta(M)$ = 10 GeV. The two signals are on arbitrary scale,
     and all previous cuts have been applied (cf. Tabs. \ref{tab:cutsflow_dm34} and  \ref{tab:cutsflow_dm10}). Events between the vertical
     lines are accepted.}
    \label{fig:thetaacop_dm34_dm10}
  \end{figure}

\begin{figure}[htbp]
    \centering
    \subcaptionbox{}{\includegraphics [width=0.45\textwidth]{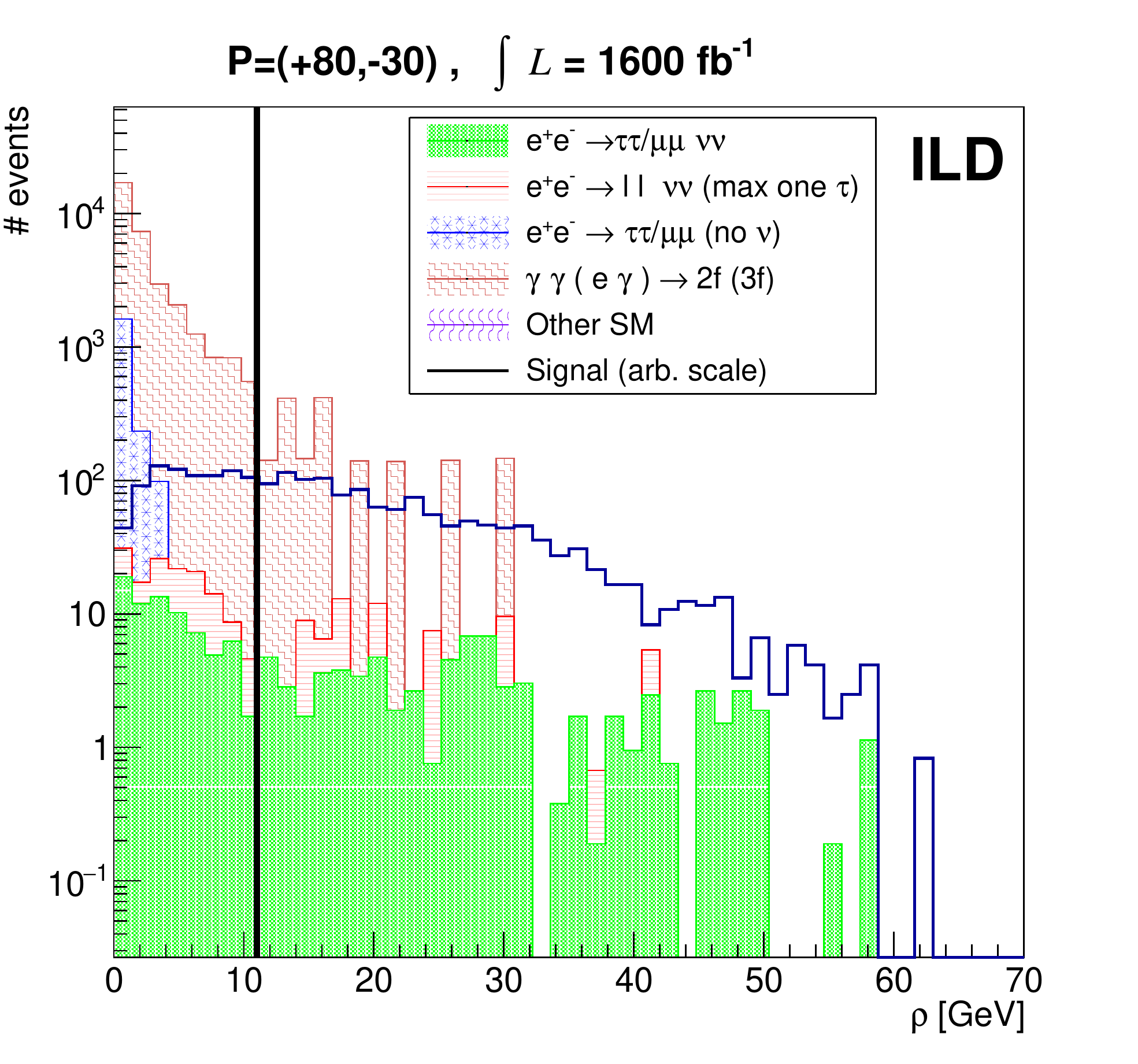}}
     \subcaptionbox{}{ \includegraphics [width=0.45\textwidth]{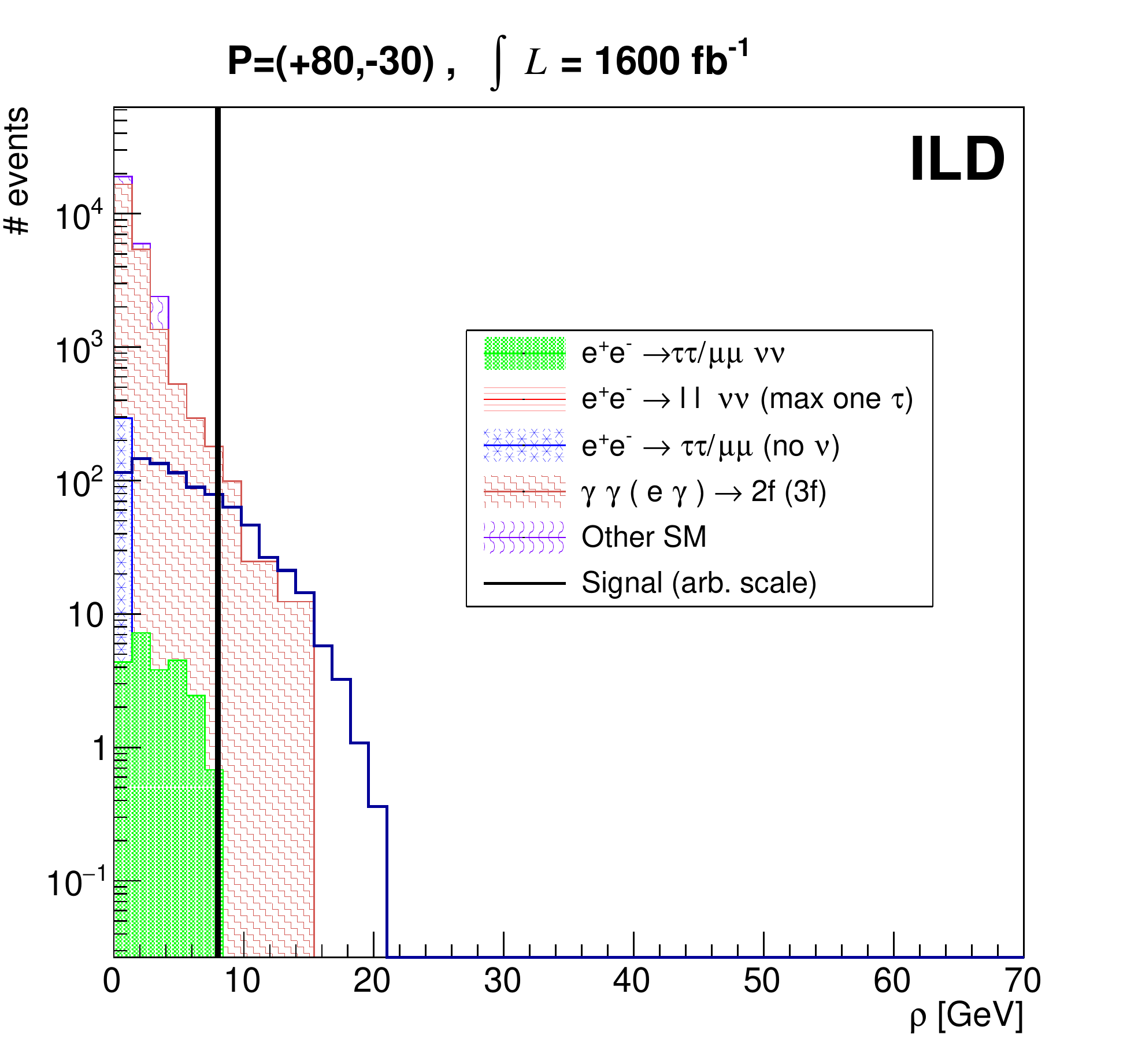}}
   \caption{
     Distributions of the variable $\rho$, described in the text, for
     (a) M$_{\widetilde{\tau}}$ = 230 GeV $\Delta(M)$ = 34 GeV,
     and (b)  M$_{\widetilde{\tau}}$ = 245 GeV $\Delta(M)$ = 10 GeV. The two signals are on arbitrary scale,
     and all previous cuts have been applied (cf. Tabs. \ref{tab:cutsflow_dm34} and  \ref{tab:cutsflow_dm10}). Events above the vertical
     lines are accepted.}
    \label{fig:rho_dm34_dm10}
  \end{figure}

  A second group of cuts is based on those properties that the
  $\widetilde{\tau}$-events {\it might} have, but will {\it rarely} be
    \begin{figure}[htbp]
    \centering
    \subcaptionbox{}{\includegraphics [width=0.45\textwidth]{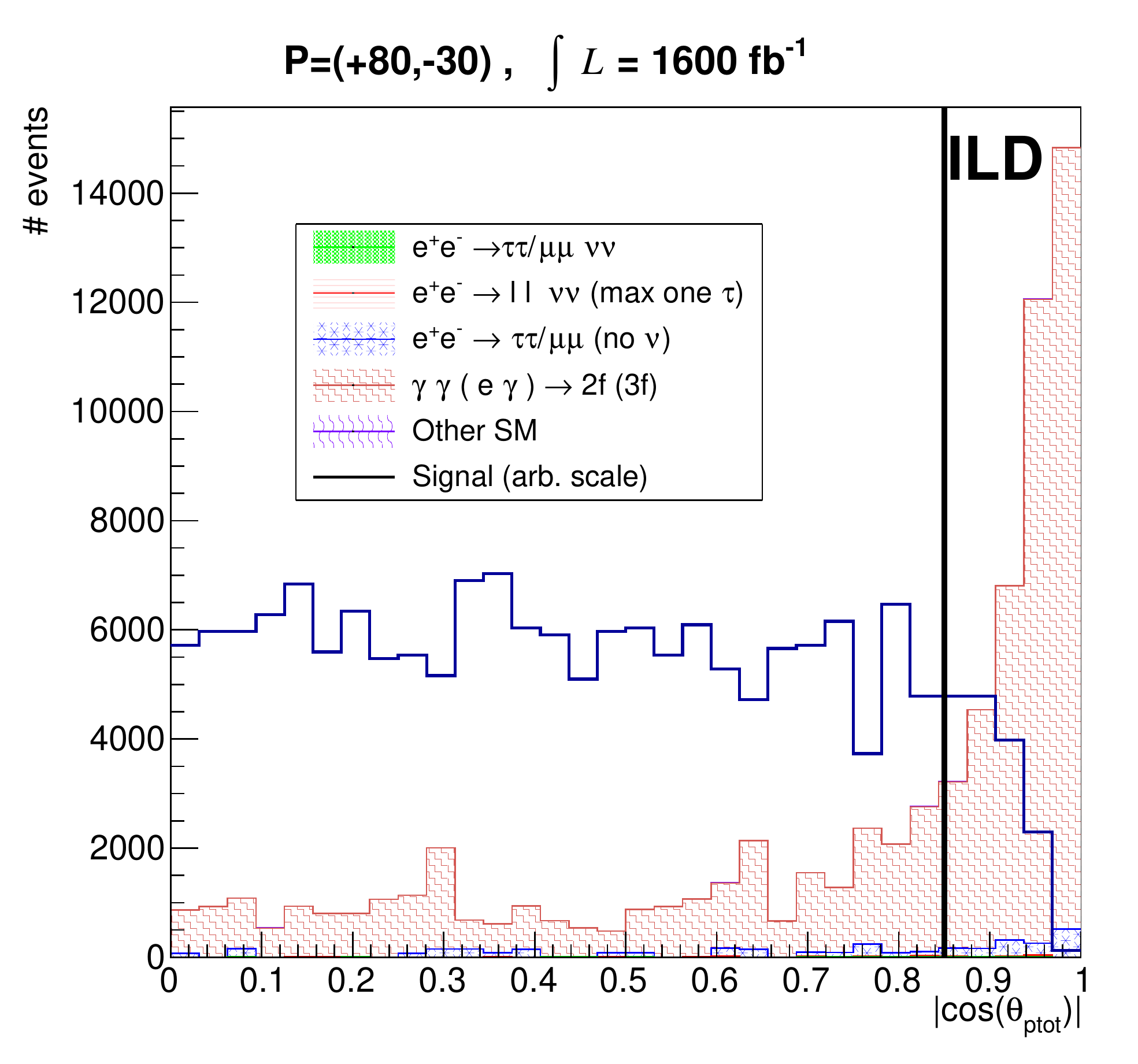}}
     \subcaptionbox{}{ \includegraphics [width=0.45\textwidth]{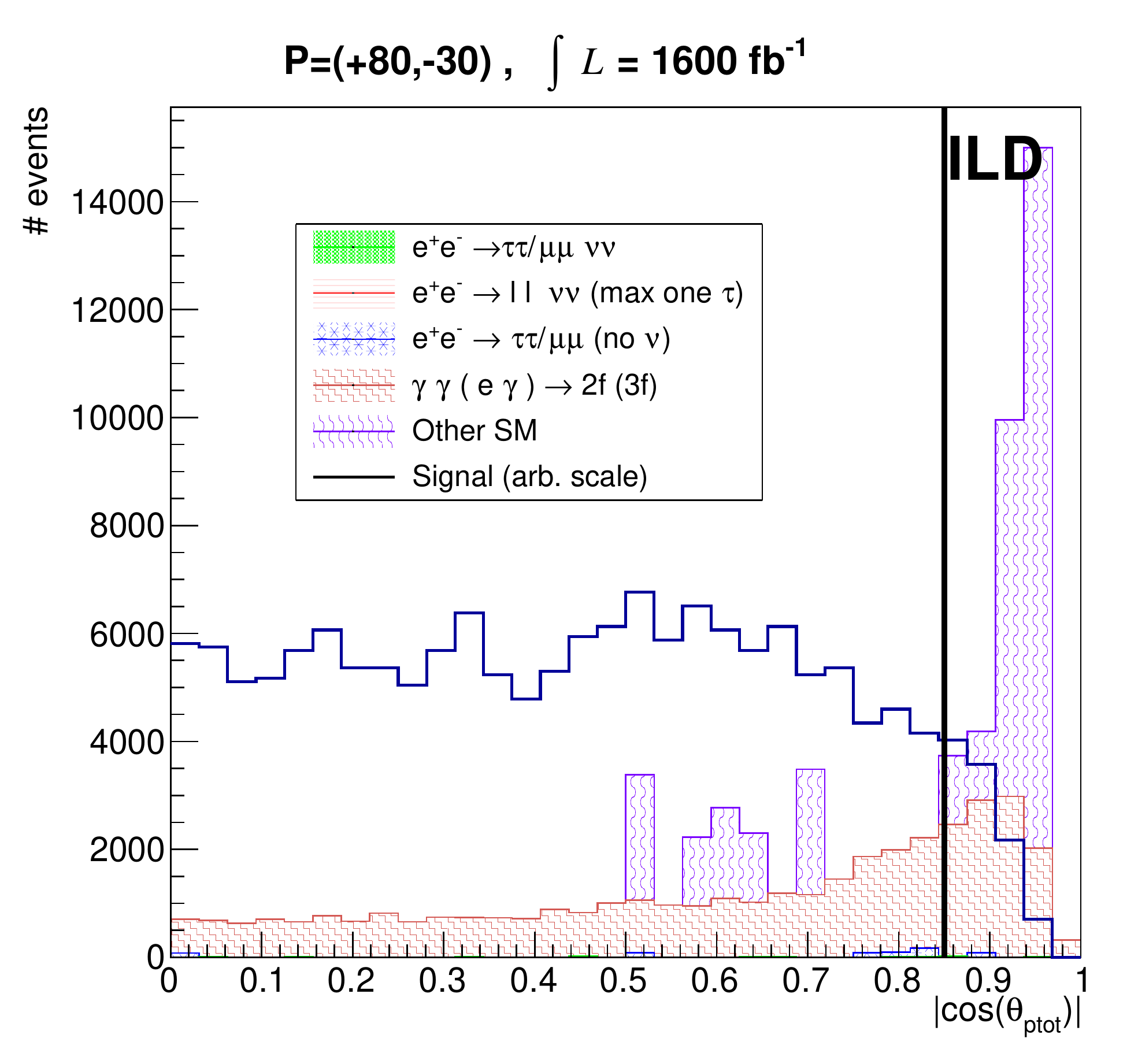}}
   \caption{
     Distributions of the polar angle of the vectorial sum of the two jet momenta for
     (a) M$_{\widetilde{\tau}}$ = 230 GeV $\Delta(M)$ = 34 GeV,
     and (b)  M$_{\widetilde{\tau}}$ = 245 GeV $\Delta(M)$ = 10 GeV. The two signals are on arbitrary scale,
     and all previous cuts have been applied (cf. Tabs. \ref{tab:cutsflow_dm34} and  \ref{tab:cutsflow_dm10}). Events below the vertical
     lines are accepted.}
    \label{fig:costh_ptot_dm34_dm10}
  \end{figure}
  present in background events. As already pointed out, the  $\widetilde{\tau}$'s are
  scalars, and therefore isotropically produced, while the backgrounds are
  either fermions or vector bosons, and tend to be produced at small angles to the beam
  axis. This allows to impose cuts requiring events to have high missing transverse
  momentum ($P_{T miss}$), and large acoplanarity $\theta_{acop}$.
  The distributions for these two quantities are shown in Figs. \ref{fig:ptmiss_dm34_dm10}
  and \ref{fig:thetaacop_dm34_dm10}, for two model-points as they are after applying all previous cuts,
  applicable to the model-points in question.
  The total seen momentum, $\sum \bar{p}$, tends to be in an almost random direction
  in signal events, while it tends to point close to the beam-axis for most
  backgrounds. Therefore a cut on the direction of the total momentum, $|\cos(\theta_{\sum \bar{p}})|<0.85$, was
  imposed, see Fig.\ref{fig:costh_ptot_dm34_dm10}. 
  In addition, a cut on
   the variable $\rho$ is imposed.
  This variable is  calculated by first projecting the event on the x-y
  (transverse) plane, and calculating the thrust axis in that plane. $\rho$
  is then the transverse momentum (in the plane) with respect to the
  thrust axis.
  The cut in $\rho$ helps to reject events with two $\tau$'s back-to-back in the transverse
  projection with the {\it visible} part of the decay of one of the $\tau$'s in the direction of its parent,
  while the other $\tau$ decays with the {\it invisible} $\nu$ closely aligned with the direction of its parent.
  These events fake the signal topology,
  having both a large missing transverse momentum and high acoplanarity.
  However, they would have
  a small value of $\rho$. The distributions of $\rho$ for signal and background, at this point in the cut chain,
  are shown in Fig. \ref{fig:rho_dm34_dm10}, for two model points.
  The values at which the cuts described in this paragraph 
  are set depends on 
 the mass difference
  between the $\widetilde{\tau}$ and the LSP (except the cut on $\theta_{\sum \bar{p}}$), and are detailed in Tab. \ref{tab:cuts}.
  \begin{table}[htbp]
    \centering
    \caption{
       Cuts depending on the mass difference between the  $\widetilde{\tau}$ and the LSP.
      The specific cuts applied for $\Delta(M)$ = 2 GeV are explained in Sect. \ref{sec:lowdmcuts}.}
    \label{tab:cuts}
    \begin{tabular}{lccc}
      \hline\hline
      \addlinespace[2pt]
      $\Delta(M)$ &  $P_{T miss}$  & $\rho$ & $\theta_{acop}$     \\[2pt]
      [GeV]       &  [GeV]        &  [GeV] & [Rad]              \\[2pt]
      \hline
      \addlinespace[2pt]
      34          & $>$ 11        & $>$ 11 & $\in [0.2 , 3.1 ] $ \\
      10          & $>$ 4         & $>$  8 & $\in [0.2 , 2.0 ] $ \\
       3          & $>$ 2         & $>$  3 & $\in [0.2 , 2.3 ] $ \\
       2          &   -           & no cut & $\in [0.2 , 2.3 ] $ \\
      \hline
      \addlinespace[2pt]
    \end{tabular}
  \end{table}
Cutting in these properties has a certain
  cost in efficiency but improves the signal-to-background ratio.
   
\begin{figure}[htbp]
    \centering
    \subcaptionbox{}{\includegraphics [width=0.45\textwidth]{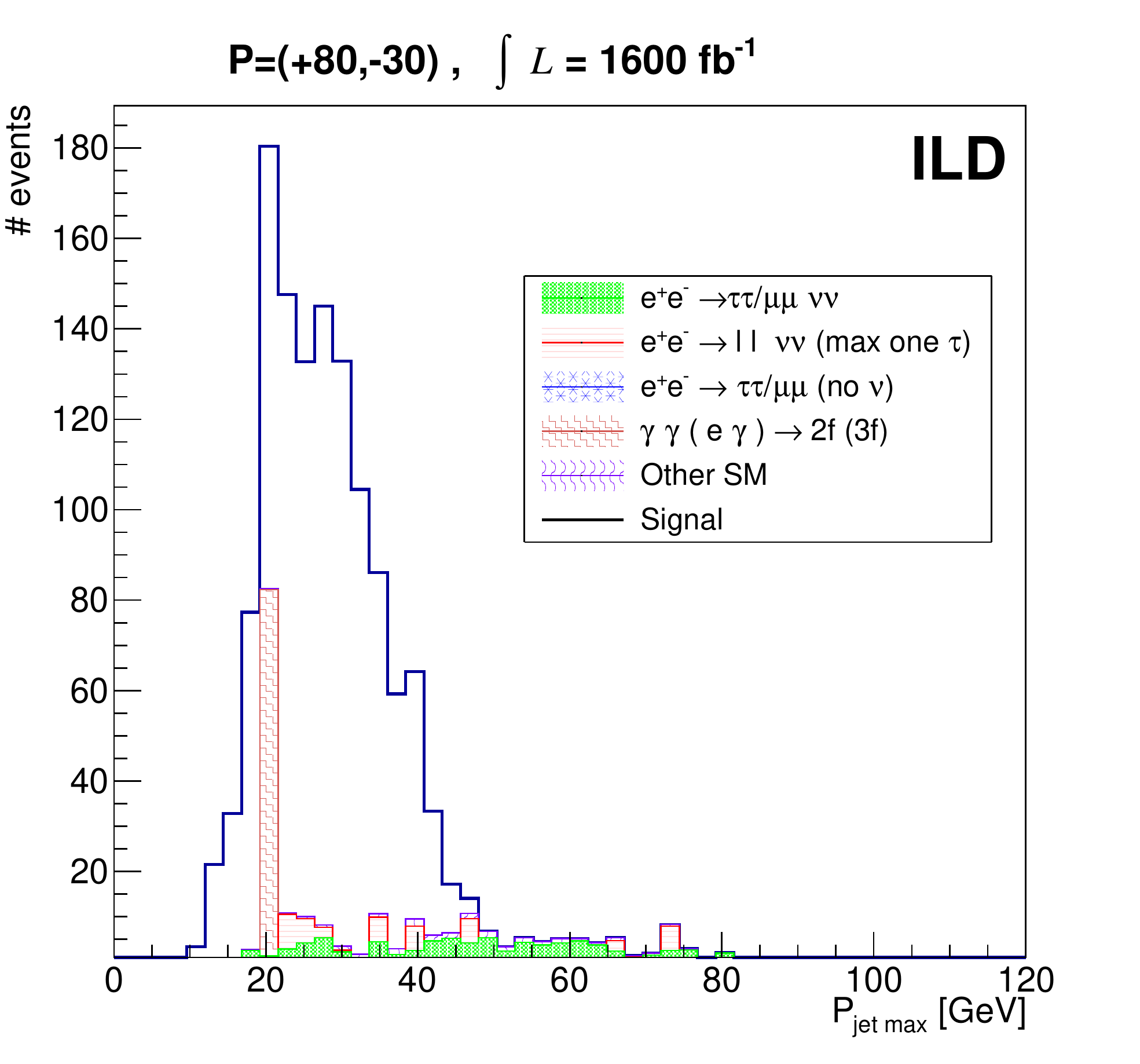}}
   \caption{
     Distribution of the higher of the two jet momenta after all cuts (except the one on $p_{jet~max}$ itself)
     corresponding to the model point M$_{\widetilde{\tau}}$ = 230 GeV, $\Delta(M)$ = 34 GeV.
     The maximum $P_{jet}$ possible at this model point is 47.63 GeV (Eq. \ref{eq:pmax}), and is indeed the endpoint of
     the signal distribution.
     (Contrary to other figures in this section, the signal is stacked on top of the background,
     and corresponds to the same integrated luminosity as the background.)}
    \label{fig:pjetmaxend_dm34}
  \end{figure}

  As a final cut it is noted that sizeable energy
  detected at low angles to the beam is rare in signal events, but common in many backgrounds.
  Events with more than 2 GeV detected at angles lower than 20
  degrees to the beam axis are therefore rejected.
  This cut is however not useful for small mass differences.
  After applying these cuts the main sources of remaining background are $WW$ events with each
  $W$ decaying to $\tau\nu$ and events with four fermions in the final state coming from
  $\gamma\gamma$ interactions, mostly $\tau\tau$ events.
  As an illustration of the expected signal after all cuts, Fig \ref{fig:pjetmaxend_dm34} shows the
  distribution of the higher of the two jet-momenta  ($p_{jet~max}$) for the model-point
  with  M$_{\widetilde{\tau}}$ = 230 GeV and $\Delta(M)$ = 34 GeV, when all described cuts except the one
  on $p_{jet~max}$ has been applied. In this figure, the signal and background
  distributions correspond to the same 1600 fb$^{-1}$ of integrated luminosity.
 
\begin{table}[htbp]
    \centering
    \caption{Flow table M$_{\widetilde{\tau}}$ = 230 GeV $\Delta(M)$ = 34 GeV. The numbers correspond to an integrated luminosity of 1.6 ab$^{-1}$ and
    polarisation $\mathcal{P}_{+-}$.}
    \label{tab:cutsflow_dm34}
    \begin{tabular}{lS[table-format=4.1]S[table-format=4.1]S[table-format=4.1]S[table-format=4.1]S[table-format=4.1]S[table-format=4.1]}
      \hline\hline
      \addlinespace[2pt]
      Cut &\alc{c}{Signal}& \alc{c}{$e^+e^- \rightarrow $}     &\alc{c}{$e^+e^- \rightarrow $}  &\alc{c}{$e^+e^- \rightarrow $} &\alc{c}{$\gamma\gamma (\gamma e) \rightarrow $}&\alc{c}{All other}  \\[2pt]
          &               & \alc{c}{$\tau\tau/\mu\mu ~\nu\nu$} &\alc{c}{$ l l ~\nu\nu$}        & \alc{c}{$\tau\tau/\mu\mu $}   &\alc{c}{2(3) fermions}                         &                    \\[2pt]
          &               &                                    &\alc{c}{($\ge 1~l$ not $\tau$)}& \alc{c}{(no $\nu$'s)}         &                                               &                    \\[2pt]
      \hline
      \addlinespace[1pt]
      no cut                             &  8292.8  &  134100 &  533900 & \alc{l}{4.377 \tento{6}}&\alc{l}{1234 \tento{6}} & \alc{l}{2616 \tento{6}}\\
      BeamCal veto                       &  8284.5  &  133966 &  532800 & \alc{l}{4.368 \tento{6}}&\alc{l}{1086 \tento{6}} & \alc{l}{2603 \tento{6}}\\
      2 $\le N_{charged} \le$ 10           &  8219.0  &  108900 &  403100 & \alc{l}{3.542 \tento{6}}&\alc{l}{1024 \tento{6}} & \alc{l}{1720 \tento{6} }\\
      $ q \cos{ \theta_{jet}} > -1$       &  7179.1  &   70720 &  216800 & \alc{l}{2.671 \tento{6}}&\alc{l}{841.6 \tento{6}}& \alc{l}{1504 \tento{6} } \\
      $|M_{vis} - M_Z| > 4$               &  7178.2  &   66820 &  207200 & \alc{l}{2.552  \tento{6}}&\alc{l}{826.8 \tento{6}}& \alc{l}{1500 \tento{6} }\\
      $E_{miss} >2 M_{LSP}$                &  7177.4  &   19620 &   14570 &                 365493.0 &\alc{l}{238.0 \tento{6}}& \alc{l}{1399 \tento{6} }\\
      $M_{vis} < \sqrt{s} - 2 M_{LSP}$    &  7103.6   &   17990 &   12490 &                 334886.7 &\alc{l}{235.5 \tento{6}}& \alc{l}{1398 \tento{6} }\\
      $\tau$ Id                         &  2953.9   &    1250 &   451.2 &                  29088.3 &\alc{l}{23.96 \tento{6}}&  \alc{l}{166.9\tento{6} }\\
      $P_{jet} < P_{max}$                 &  2953.9   &    1003 &   268.3 &                  23963.2 &\alc{l}{23.55 \tento{6}}&  \alc{l}{165.6\tento{6} }\\
      $P_{T miss}$ > 11 GeV               &  2478.7   &   242.5 &   175.6 &                     3481 &                  83970 &                     54.7 \\
      $\theta_{acop} \in$  [0.2, 3.1]     &  2293.8  &    208.6 &  144.9 &                      3075&                   69190 &                    52.0 \\
      $|\cos(\theta_{\sum \bar{p}})|<0.85$  &  2125.4  &   147.1 &   113.6 &                      1876&                   32520 &                    39.8 \\
      $\rho$ >  11                       &  1317.7  &    60.3 &    39.3 &                      0.0 &                    1923 &                    19.3 \\
      $E_{<20^\circ} < 2$~GeV              &  1015.9  &    39.6 &    27.2 &                      0.0 &                    81.0 &                    12.1 \\
      \hline
      \addlinespace[2pt]
    \end{tabular}
  \end{table}

\begin{table}[htbp]
    \centering
    \caption{Flow table M$_{\widetilde{\tau}}$ = 245 GeV $\Delta(M)$ = 10 GeV. The numbers correspond to an integrated luminosity of 1.6 ab$^{-1}$ and
    polarisation $\mathcal{P}_{+-}$.}
    \label{tab:cutsflow_dm10}
    \begin{tabular}{lS[table-format=4.1]S[table-format=4.1]S[table-format=4.1]S[table-format=4.1]S[table-format=4.1]S[table-format=4.1]}
      \hline\hline
      \addlinespace[2pt]
      Cut &\alc{c}{Signal}& \alc{c}{$e^+e^- \rightarrow $}     &\alc{c}{$e^+e^- \rightarrow $}  &\alc{c}{$e^+e^- \rightarrow $} &\alc{c}{$\gamma\gamma (\gamma e) \rightarrow $}&\alc{c}{All other}  \\[2pt]
          &               & \alc{c}{$\tau\tau/\mu\mu ~\nu\nu$} &\alc{c}{$ l l ~\nu\nu$}        & \alc{c}{$\tau\tau/\mu\mu $}   &\alc{c}{2(3) fermions}                         &                    \\[2pt]
          &               &                                    &\alc{c}{($\ge 1~l$ not $\tau$)}& \alc{c}{(no $\nu$'s)}         &                                               &                    \\[2pt]
     \hline
      \addlinespace[1pt]
      no cut                             &  672.8   &  134100 &  533900 & \alc{l}{4.377 \tento{6}}&\alc{l}{1234 \tento{6}} & \alc{l}{2616 \tento{6}}\\
      BeamCal veto                       &  672.1  &  133966 &  532800 & \alc{l}{4.368 \tento{6}}&\alc{l}{1086 \tento{6}} & \alc{l}{2603 \tento{6}}\\
      2 $\le N_{charged} \le$ 10         & 664.3 &  106400 &  403100 & \alc{l}{3.542 \tento{6}}&\alc{l}{1024 \tento{6}}&\alc{l}{1720 \tento{6}}\\
      $ q \cos{ \theta_{jet}} > -1$     & 585.2 &   70720 &  216800 & \alc{l}{2.671 \tento{6}}&\alc{l}{841.7 \tento{6}}&\alc{l}{1504 \tento{6}}\\
      $|M_{vis} - M_Z| > 4$             & 585.2 &   66820 &  207200 & \alc{l}{2.552 \tento{6}}&\alc{l}{826.8 \tento{6}}&\alc{l}{1500 \tento{6}} \\
      $E_{miss} >2 M_{LSP}$              & 585.1 &    6218 &   886.6 &                  120800 &\alc{l}{179.0 \tento{6}}&\alc{l}{1043 \tento{6}}\\
      $M_{vis} < \sqrt{s} - 2 M_{LSP}$   & 583.9 &    5669 &   758.0 &                  109100 &\alc{l}{172.7 \tento{6}}&\alc{l}{969.9 \tento{6}}\\
      $\tau$ Id                        & 230.6 &   484.0 &    56.3 &                   13020 &\alc{l}{1.010 \tento{6}}&\alc{l}{127.4 \tento{6}}\\
      $P_{jet} < P_{max}$                & 229.9 &   444.2 &    38.2 &                   11150 &                822400  &\alc{l}{108.0 \tento{6}}\\
      $P_{T miss}$ > 4 GeV               & 177.4 &   33.4 &      5.1 &                   435.0 &                  36210 &       84660 \\
      $\theta_{acop} \in$  [0.2, 2.0]    & 164.6 &   30.0 &     5.1 &                   299.6 &                   32310 &       71650 \\
      $|\cos(\theta_{\sum \bar{p}})|<0.85$ & 156.4 &   22.7 &     0.0 &                   290.0 &                  24220  &       20290 \\
      $\rho$ >  8                       &  55.6 &    2.7 &     0.0 &                     0.0 &                   798.6 &         0.5 \\
      $E_{<20^\circ} < 2$~GeV             &  55.6 &    2.7 &     0.0 &                     0.0 &                   610.2 &          0.5 \\
      \hline
      \addlinespace[2pt]
    \end{tabular}
  \end{table}

    The polarisation plays an important role in the capability of excluding/discovering
  the different regions of the SUSY space.
  Table \ref{tab:polarisations} shows the number of signal and background events for a specific model point for
  the two main ILC running polarisations and for unpolarised beams.
  The difference in the number of signal events arises from the dependence of the
  cross section on the polarisation, as well as from the effect on selection efficiency due to the 
  polarisation of the $\tau$ coming from the $\widetilde{\tau}$.
  The dependence of the
  cross section on the polarisation is the main factor for the difference in $WW$ events, $e^+e^- \rightarrow\tau\nu\tau\nu$.
  One can see that the signal-to-background ratio is clearly enhanced for the case of mainly right-handed electrons,
  left-handed positrons.
  Taking the definition of exclusion at 95$\%$ CL as $N_\sigma > 2$ (cf. Eq. \ref{eq:LR})\footnote{Note that Eq. \ref{eq:LR} trivially states that $N_\sigma = S/\sqrt{S+B}$ in the case that
    there is only one sample.},
  it can be seen that combining the two polarisation samples with the LR statistic does strengthen the limit,
  even though the  $\mathcal{P}_{-+}$ sample gives a much weaker limit than the  $\mathcal{P}_{+-}$ one. 
  It is also shown that unpolarised beams would allow neither exclusion nor discovery, even though the sample corresponds to
  the same integrated luminosity as the sum of the two polarised samples.
  This is not only because of the possibility to combine samples with different beam polarisations in an optimal way
  is not available for unpolarised beams; it is also because the ILC has {\it both} beams polarised, meaning that {\it more
  than half} of the collisions are between opposite polarised particles, which is necessary to allow for s-channel processes,
  such as $\widetilde{\tau}$ pair production.
  Polarisation is not only important in the enhancement of the signal over background but also
  plays an important role in the parameter determination.
  \begin{table}[htbp]
    \centering
    \caption{
      Remaining signal and background events after the application of the selection cuts for
      $M_{\widetilde{\tau}}$=245~\,GeV and mass difference with the LSP of 10~\,GeV. 
    }
    \label{tab:polarisations}

    \begin{tabular}{lS[table-format=4.1]S[table-format=4.1]S[table-format=4.1]S[table-format=4.1]S[table-format=4.1]}
      \hline\hline
      \addlinespace[2pt]
      Polarisation &\alc{c}{Signal}& \alc{c}{$e^+e^- \rightarrow $}     &\alc{c}{$\gamma\gamma (\gamma e) \rightarrow $}&\alc{c}{All other}  & \alc{c}{N$_\sigma$}\\[2pt]
          &               & \alc{c}{$\tau\tau/\mu\mu ~\nu\nu$}          &\alc{c}{2(3) fermions}                         &                    & \\[2pt]
      \hline
      \addlinespace[1pt]
      $\mathcal{P}_{+-}$   &  55.6 &    2.7 & 610.2 & 0.5 & 2.1 \\
      $\mathcal{P}_{-+}$   & 36.0  & 39.6   &  1208 & 0.5 & 1.0\\
      Combined              & \alc{c}{-}   & \alc{c}{-}  & \alc{c}{-} & \alc{c}{-}  &2.4\\
      Unpolarised          & 73.9  & 34.1  & 1818 & 1.0 & 1.7\\
     \hline
      \addlinespace[2pt]
    \end{tabular}


  \end{table}
  
\subsection{ Low $\Delta(M)$ cuts \label{sec:lowdmcuts}}
  The cuts described above are suited for mass differences down to 3~\,GeV.
  When the mass difference is between 3~\,GeV and the mass of the $\tau$ \footnote{For mass differences
    below the mass of the $\tau$ the lifetime of the $\widetilde{\tau}$ increases exponentially and
    the study has to be done based on a signature of long-lived particles travelling through the
    detector.} the kinematics of the signal events is very close to that of the $\gamma\gamma$ background
  events and the described cuts are not enough for discovering/excluding the signal.
  An additional cut was done based on the Initial State Radiation photons (ISR). Events with isolated photons
  with sizeable energy were selected, allowing to extension of the limits into the region under study.
  To select these, the photon should be at above 7$^\circ$ to the beam - the lower edge of the acceptance of the
  tracking system of ILD - to avoid that electrons would be 
  mistaken for photons, and it should have an energy above 1.1 GeV.
  
  This cut is effective against the remaining $\gamma\gamma$ background because these events become candidates due
  to fake missing transverse momentum. If the presence of an ISR is requested, the incoming electron or positron that
  emitted the ISR must have recoiled against the ISR.
  Since this is a
  scattering process, not an annihilation one, the electron (positron) is still present in the final state.
  Therefore, if it is required to see a high transverse momentum ISR, the final state
  electron (positron) will have acquired a recoil transverse momentum big enough to be deflected into the BeamCal, and
  thus to have been rejected already at the pre-selection stage. On the other hand, if the ISR was emitted from
  an electron or positron that was subsequently annihilated into a $Z$, as is the case for the signal process,
  the transverse momentum of the ISR is included
  in the decay products of the $Z$, and no signal is expected in the BeamCal.

  Events containing such ISR photons, and passing all cuts except those detailed in Tab. \ref{tab:cuts}, were retained
  if $P_{T miss} > 4$ GeV and the sum of the absolute value of the jet momenta  was between 1 and 4 GeV. 
\subsection{Determining the worst scenario\label{sec:worst}}
  
\begin{figure}[htbp]
    \centering
\includegraphics [width=.6\textwidth]{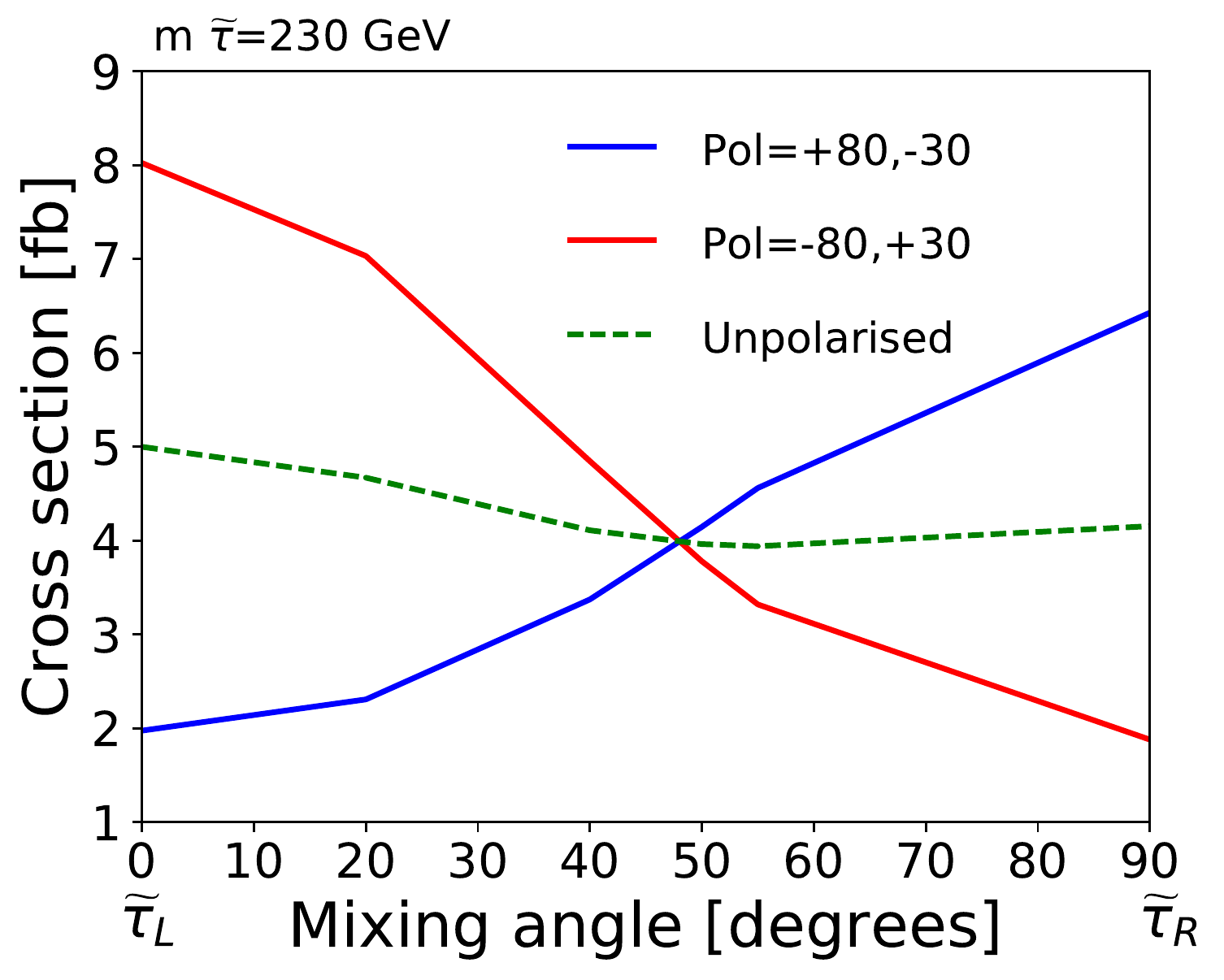}
%
    \caption{
      Cross section for $\widetilde{\tau}$ pair production as a function of the $\widetilde{\tau}$ mixing
      angle for the two main ILC polarisations. The green dotted line shows the case corresponding to the simply combination of both samples, corresponding
      to unpolarised beams for equal integrated luminosity.
      }
    \label{crosssections}
  \end{figure}
\begin{figure}[htbp]
    \centering
    \includegraphics [width=0.6\textwidth]{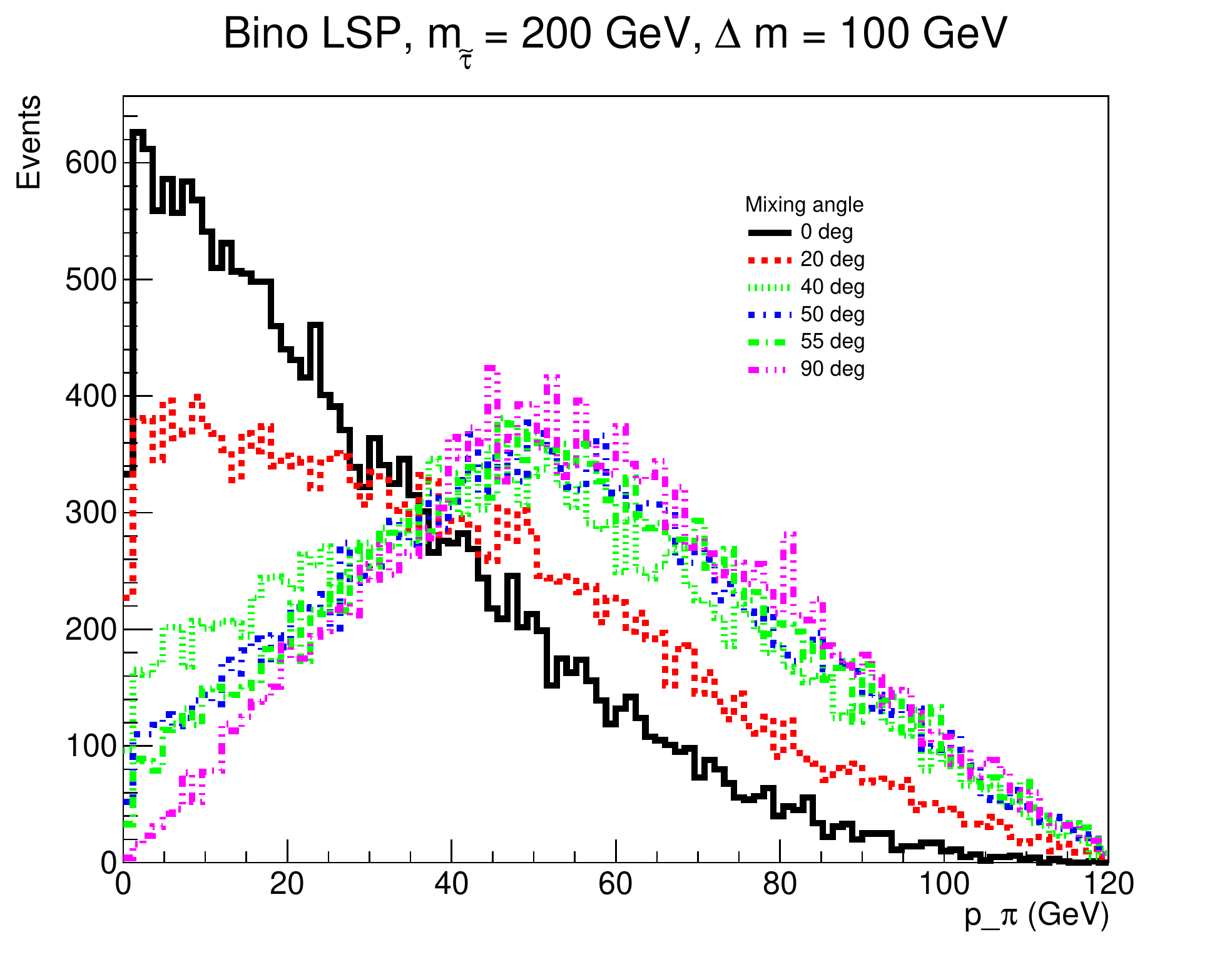}
    \caption{Momentum distribution of the pions coming from ${\tau}$-decays in the mode $\tau \rightarrow \pi^\pm \nu$ for different $\widetilde{\tau}$ mixing angles.
      The neutralino was assumed to be a pure bino.}
    \label{momentum_mixings}
  \end{figure}
\begin{figure}[htbp]
    \centering
     \subcaptionbox{}{\includegraphics [width=0.45\textwidth]{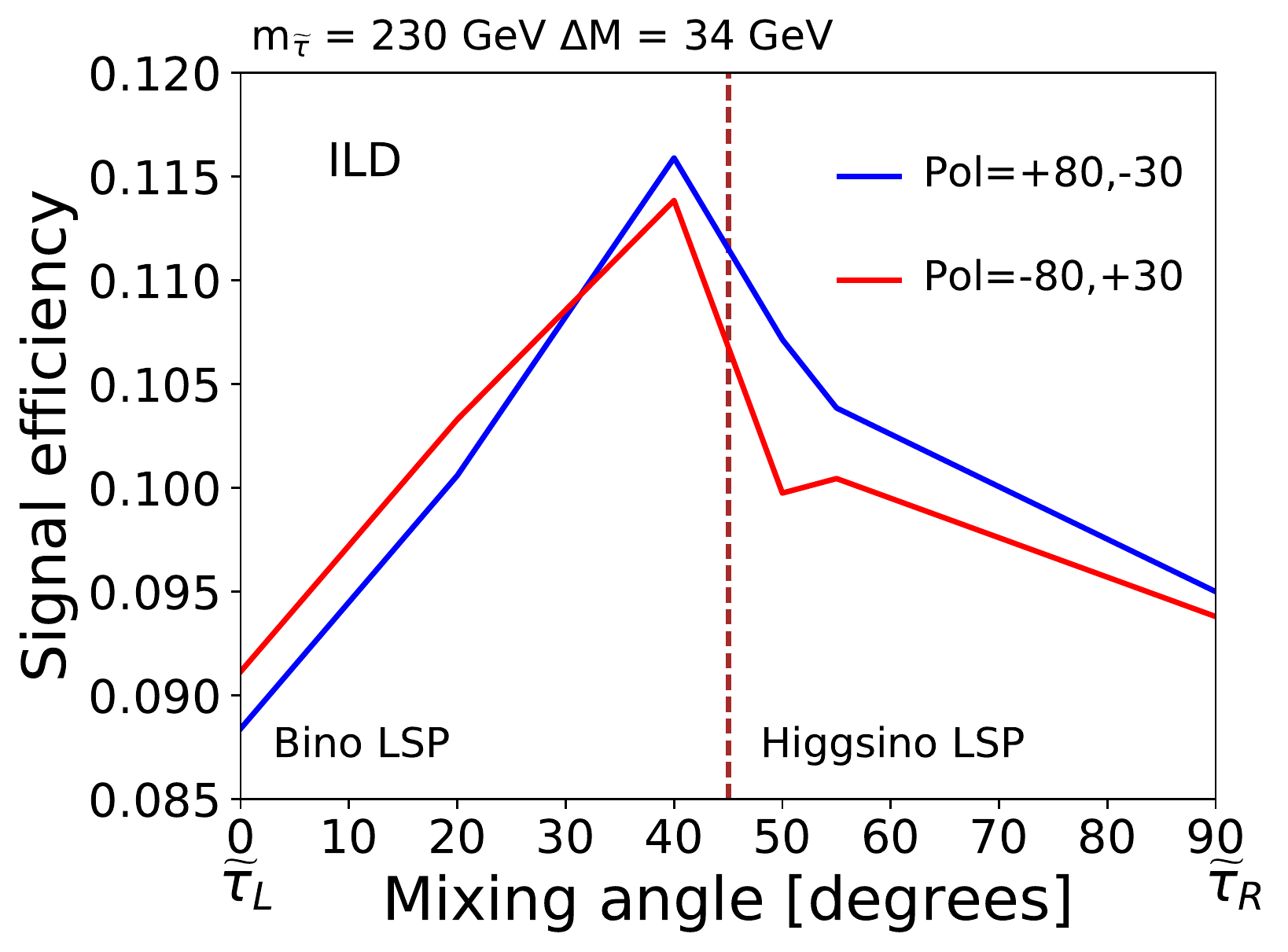}}
     \subcaptionbox{}{\includegraphics [width=0.45\textwidth]{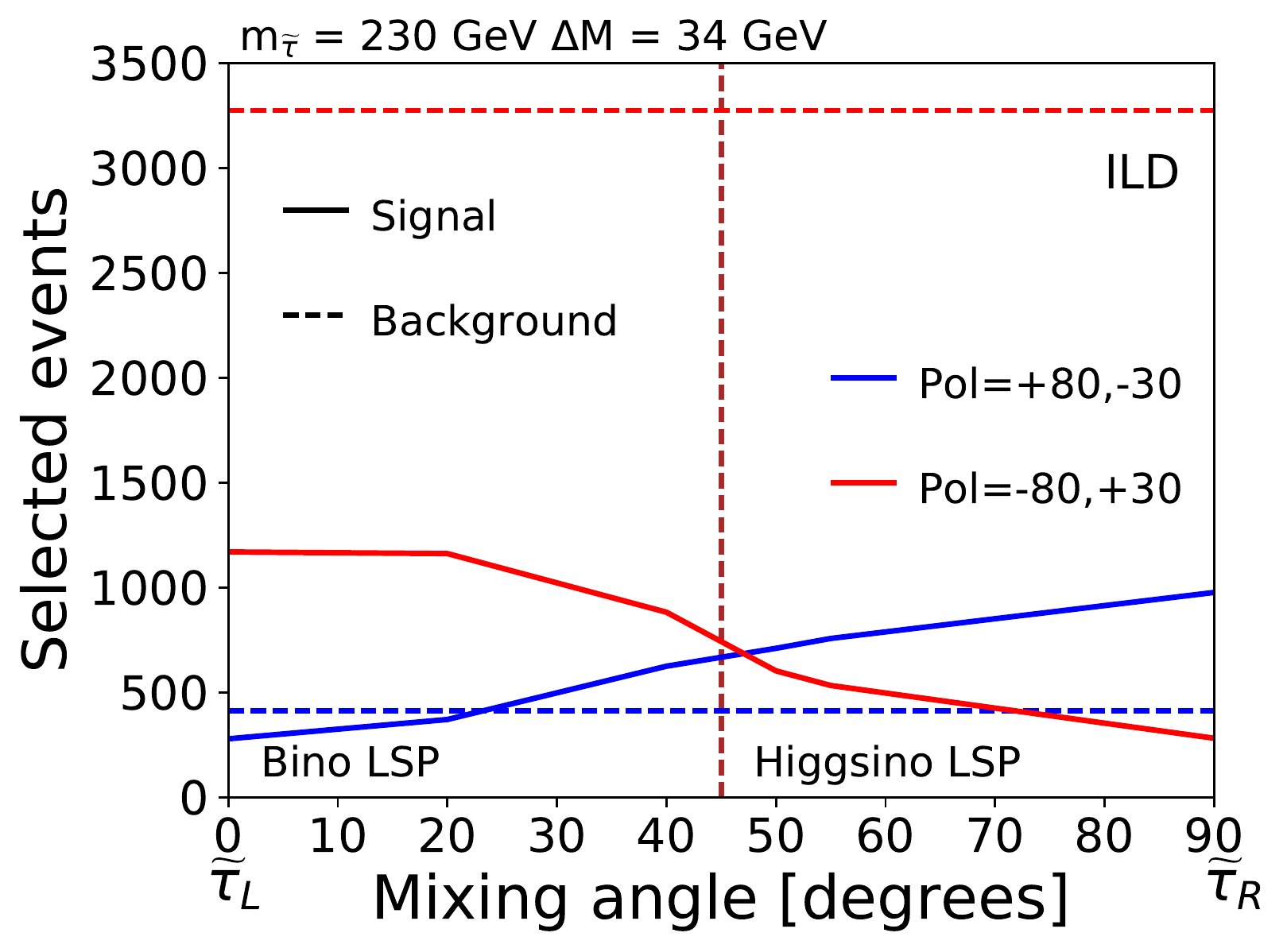}}
    \caption{(a): Signal efficiency as a function of the $\widetilde{\tau}$ mixing angle for both main ILC polarisations.
      The LSP composition was chosen giving the worst case for each mixing.
       (b):Number of signal and background selected events as a function of the $\widetilde{\tau}$ mixing angle for both main ILC polarisations. }
    \label{signal_eff_events_mixings}
  \end{figure}

The aim of this study is to evaluate the capabilities of the $\widetilde{\tau}$ search at ILC
with no loopholes.
One of the properties of the $\widetilde{\tau}$ - apart from its and the LSP masses - is
the $\widetilde{\tau}$ mixing.
This enters in two ways: The production cross-section depends on it,
as does the polarisation of the $\tau$.
The dependence of the cross-section on the $\widetilde{\tau}$ mixing angle is
shown in Fig. \ref{crosssections}, for the two beam-polarisations considered at
ILC.
Also shown is dependence for unpolarised beams, which
shows a minimum at 53$^\circ$.
The polarisation of the $\tau$ is also important, since it influences the momentum 
distribution of the ${\tau}$-decay products,
and hence the signal efficiency: if the $\tau$ is pre-dominantly left-handed, and
since there are no right-handed neutrinos in the standard model, the invisible
neutrino will tend to align with the direction of the $\tau$ and take a larger 
fraction of the momentum, and consequently the visible system will take less.
This is illustrated in Fig. ~\ref{momentum_mixings} which shows the momentum distribution 
of the pions coming from ${\tau}$-decays (in the single pion decay mode) for different
$\widetilde{\tau}$ mixing angles and a bino-LSP.
The ${\tau}$ polarisation from the $\widetilde{\tau}$ decay depends not only on the $\widetilde{\tau}$ but also on the
neutralino mixing angle. This is because of the different behaviour of the interaction between higgsinos and
binos with the $\widetilde{\tau}$. The higgsino (like the Higgs) carries weak hyper-charge, while the bino
(like the $Z$ or the photon) does not. Therefore, with a higgsino LSP, the $\widetilde{\tau}$  produces 
a ${\tau}$ with the opposite chirality with respect to that of the $\widetilde{\tau}$, while with a bino LSP, 
the produced ${\tau}$ has the same chirality.
Since the lowest efficiency is expected for left-handed $\tau$'s, we
evaluate the efficiency for the assumption that the neutralino is a pure Bino for
$\widetilde{\tau}$ mixings below 45$^\circ$ (i.e. for a $\widetilde{\tau}$ more left
than right), and for a pure higgsino LSP for mixings above  45$^\circ$.
This assures that the most difficult case is studied at any mixing.
In Fig. 
\ref{signal_eff_events_mixings}(a) 
the signal efficiency for an example
point is shown, with these assumptions.
  

\begin{figure}[htbp]
    \centering
    \subcaptionbox{}{\includegraphics [width=0.45\textwidth]{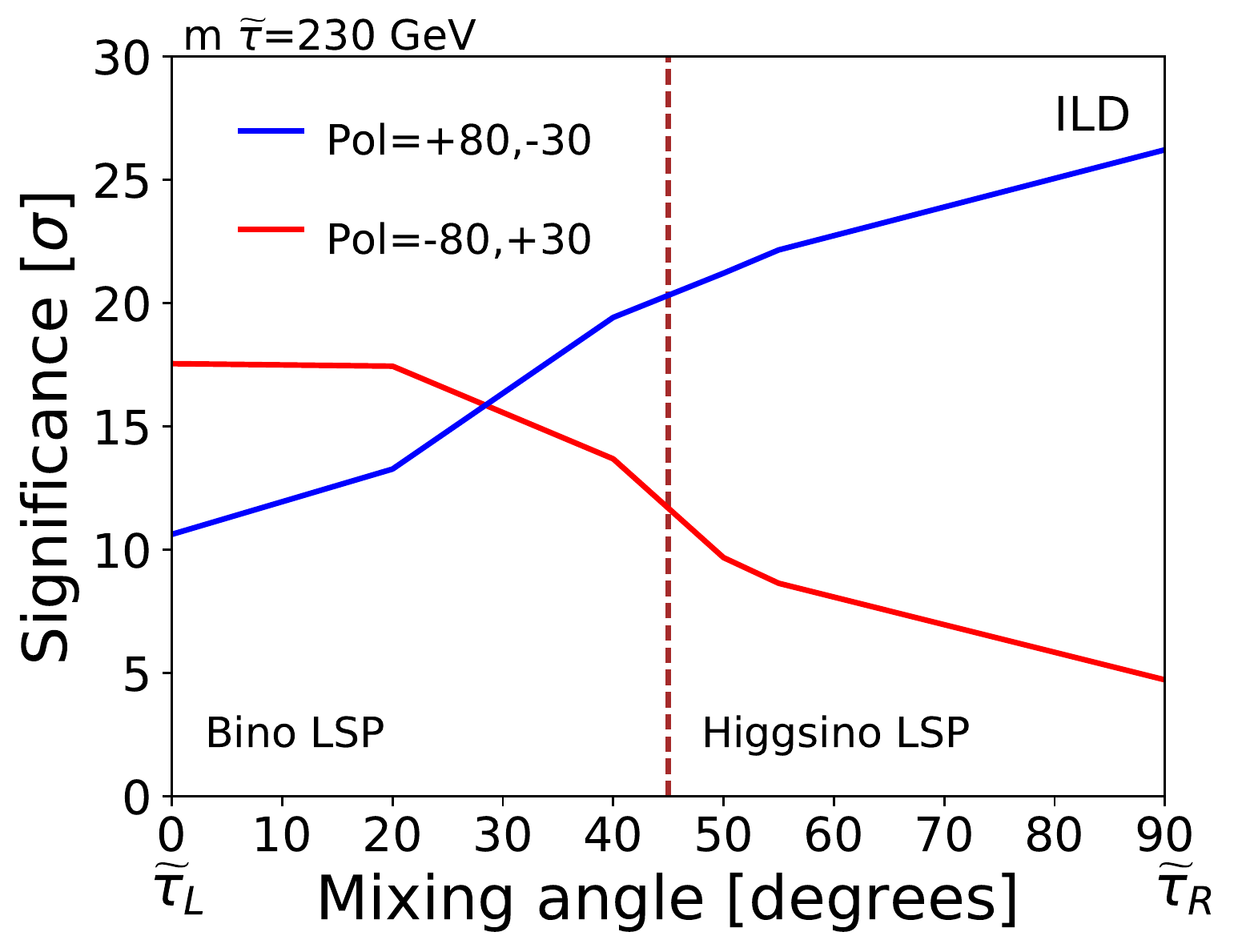}}
     \subcaptionbox{}{ \includegraphics [width=0.45\textwidth]{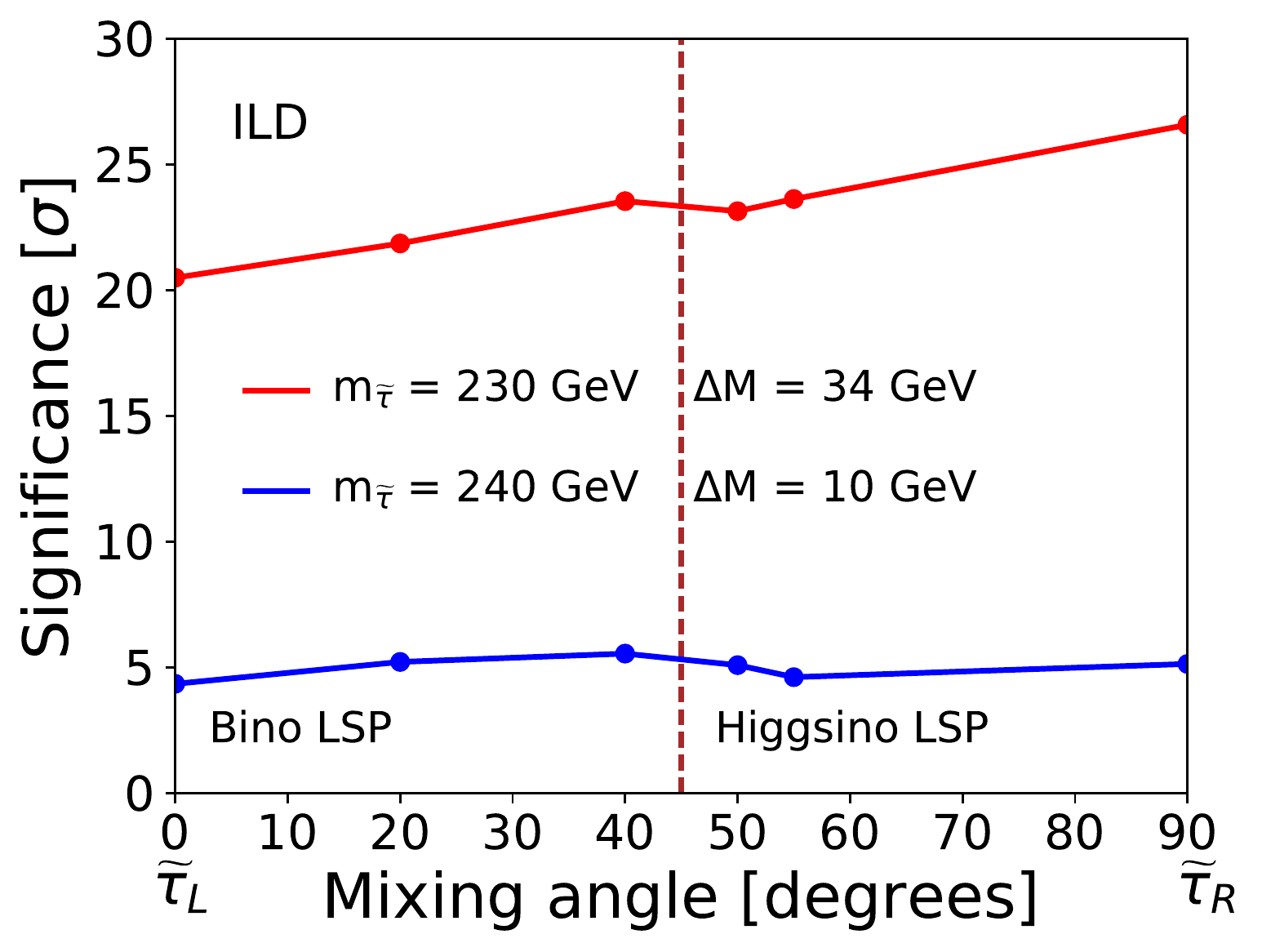}}
   \caption{(a): Signal significance as a function of the $\widetilde{\tau}$ mixing angle for both main ILC polarisations.
    (b)Signal significance weighting both polarisations using the likelihood-ratio static in the H20 ILC conditions.}
    \label{sigmas_mixings_weighted}
  \end{figure}

Combining cross sections and signal efficiencies, the number of selected signal events for each polarisation as
a function of the $\widetilde{\tau}$ mixing angle was evaluated. Figure~\ref{signal_eff_events_mixings}(b) shows the obtained
values together with the number of background events passing the selection cuts, which is different for each
beam polarisation but obviously do not have any dependence on the $\widetilde{\tau}$ mixing angle.
As can be seen, the background is much larger for the $\mathcal{P}_{-+}$ sample.
This is because the irreducible background from $e^+ e^- \rightarrow W W  \rightarrow \tau \nu \tau \nu$ is
strongly polarisation dependent.
This constitutes the final ingredient to estimate the worst-case scenario:
from these values, the signal over background significance as a function of the mixing angle was computed for
each polarisation, as shown in figure~\ref{sigmas_mixings_weighted}(a). 
Final significance values were computed adding the contribution of
both polarisations weighted by the likelihood ratio statistic.
The results are plotted in
figure~\ref{sigmas_mixings_weighted}(b). 
One can see that rather uniform sensitivity to all mixing angles is obtained, and that for the 
smallest mass differences - the ones closest to the critical region - a mixing angle around 53$^\circ$ can indeed 
be considered as the
worst one,  and validates our choice of this mixing angle being the worst case\footnote{
Note that the estimation of the worst scenario was not done with the final cuts used in
the analysis. The final cuts were further optimised taking into account the whole parameter
space. This does not affect the conclusion of this section.}.

\subsection{Effect of overlay tracks \label{sec:overlcuts}}
  
\begin{figure}[htbp]
    \centering
    \subcaptionbox{}{\includegraphics [width=0.45\textwidth]{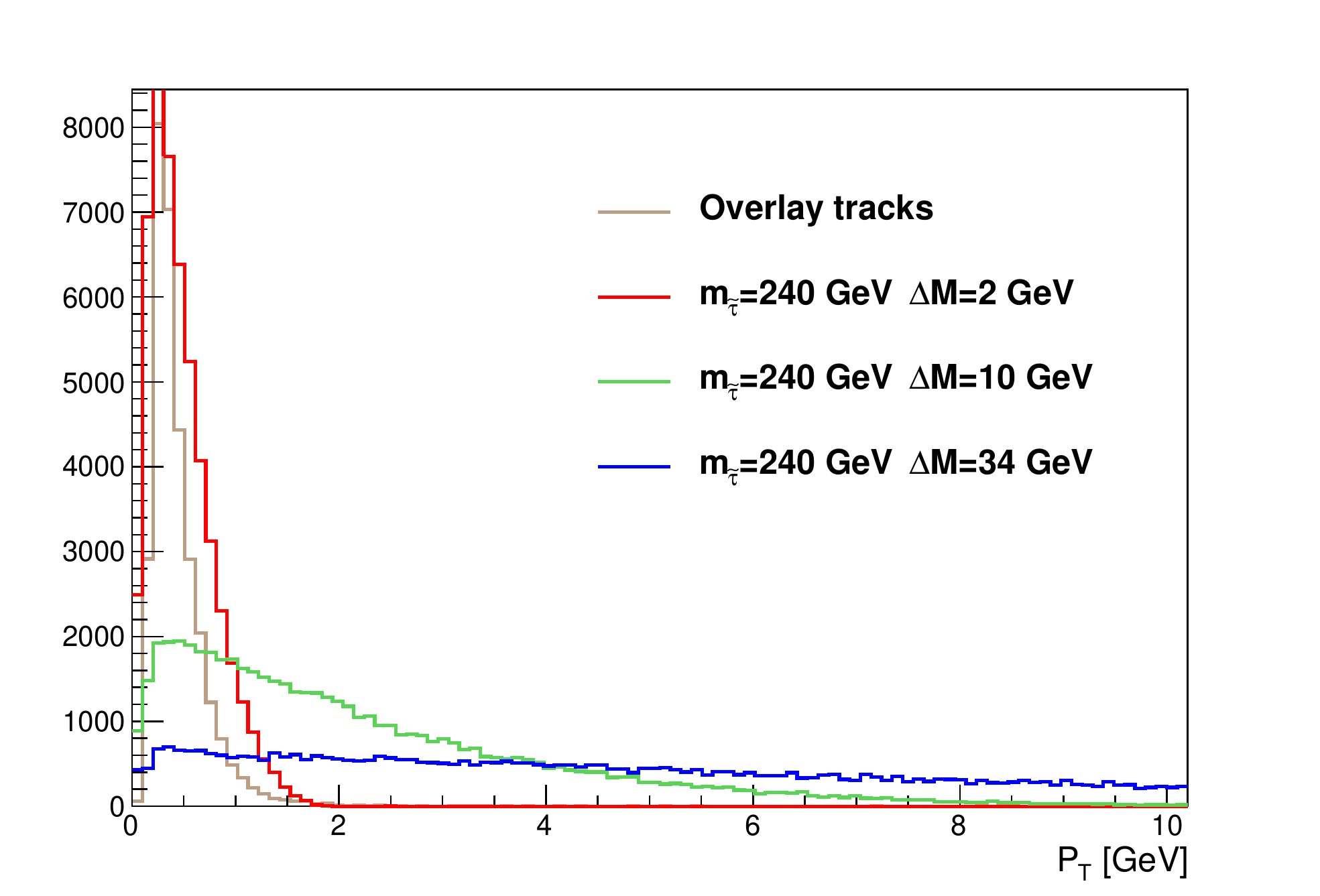}}
    \subcaptionbox{}{\includegraphics [width=0.45\textwidth]{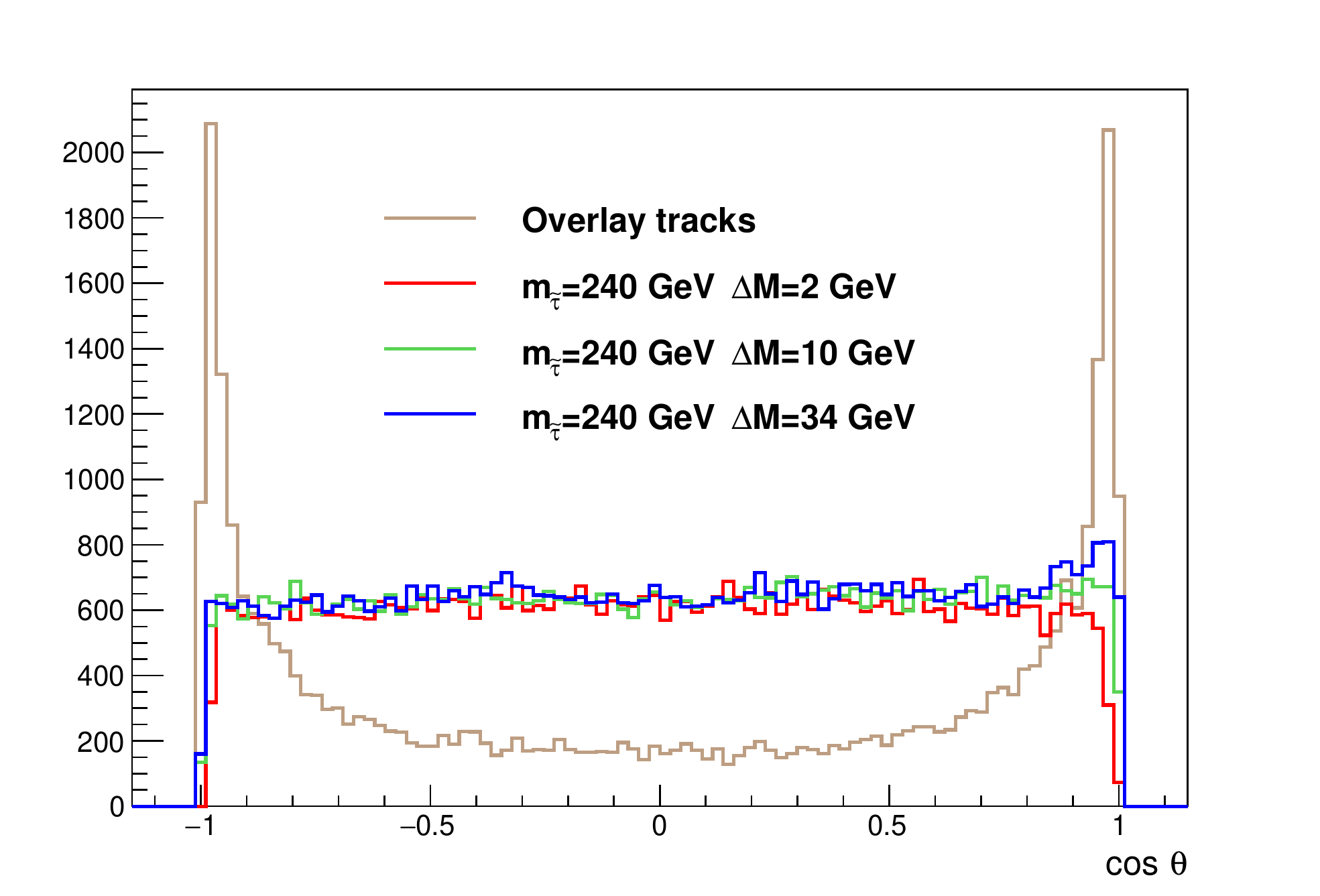}}
    \caption{(a): Transverse momentum distribution for signal and overlay tracks. Signal tracks correspond
      to different mass differences between $\widetilde{\tau}$ and LSP.
      (b): Distribution of cosine of polar angle for signal and overlay tracks. Signal tracks correspond
      to different mass differences between $\widetilde{\tau}$ and LSP.}
    \label{pt_theta_signal_overlay}
  \end{figure}

The overlay tracks can not be neglected in the $\widetilde{\tau}$ studies, since
they have similar properties to the visible tracks from the $\widetilde{\tau}$-decays in the region
with small mass differences between $\widetilde{\tau}$ and LSP.
The main
characteristics of the overlay tracks are the low transverse momentum and low angles to the
beam axis.
The tracks in low $\it{P_{T}}$ hadron events usually originates in the beam-spot,
contrary to the tracks from the $\tau$ decays.
Figure~\ref{pt_theta_signal_overlay}(a) compares transverse momentum distributions for signal and overlay tracks,
and it can be seen that while the overlay tracks can easily be removed by a cut in this quantity for
large to moderate mass-differences, this is not possibly for the smallest ones.
Figure~\ref{pt_theta_signal_overlay}(b) makes the same comparison for the cosine of the polar angle of the
tracks, showing that in this variable, the difference between overlay and signal tracks does not depend
on the mass-difference.

\begin{figure}[htbp]
    \centering
    \subcaptionbox{overlay}{\includegraphics [width=0.45\textwidth]{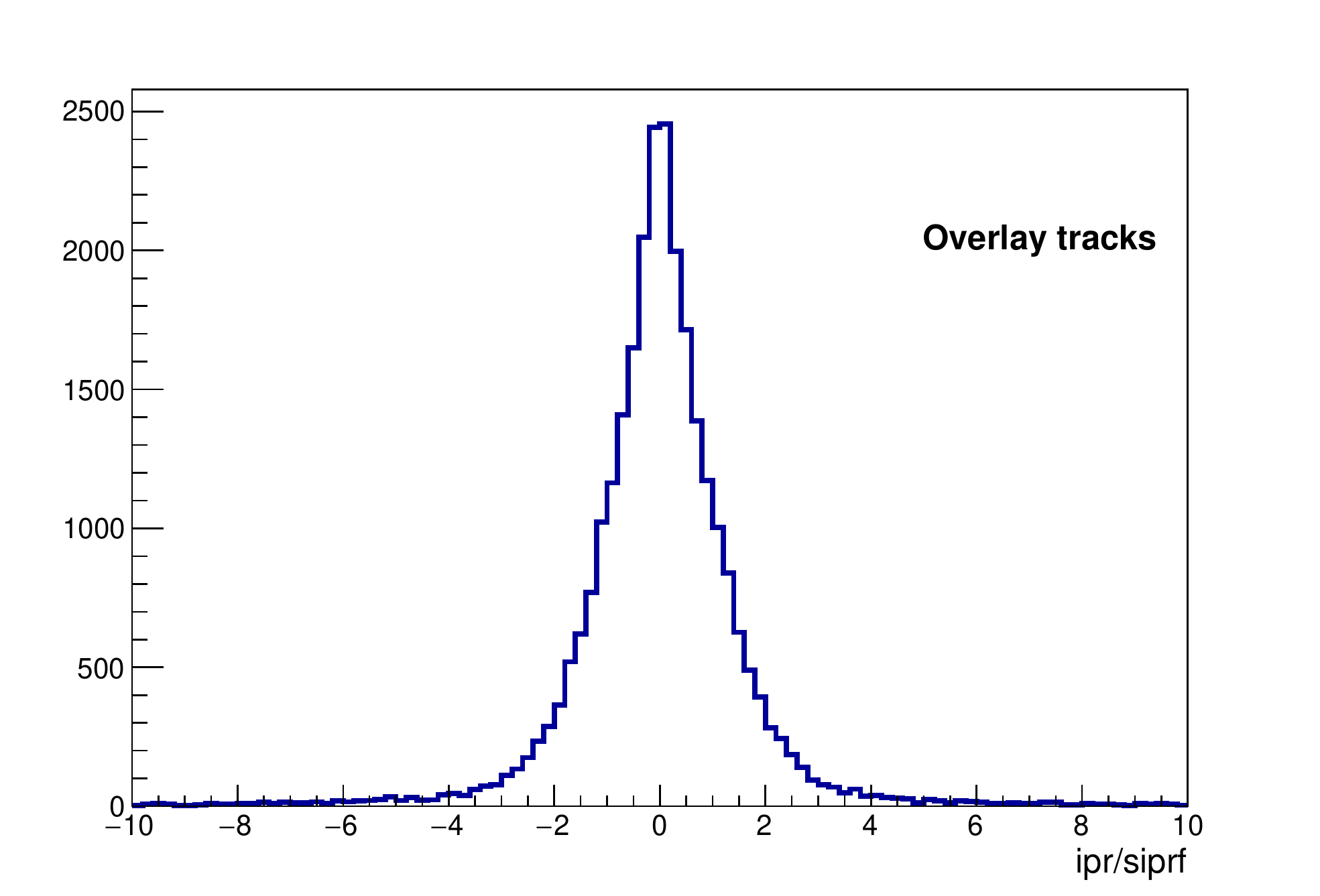}}
    \subcaptionbox{$\Delta(M)$ = 2 GeV}{\includegraphics [width=0.45\textwidth]{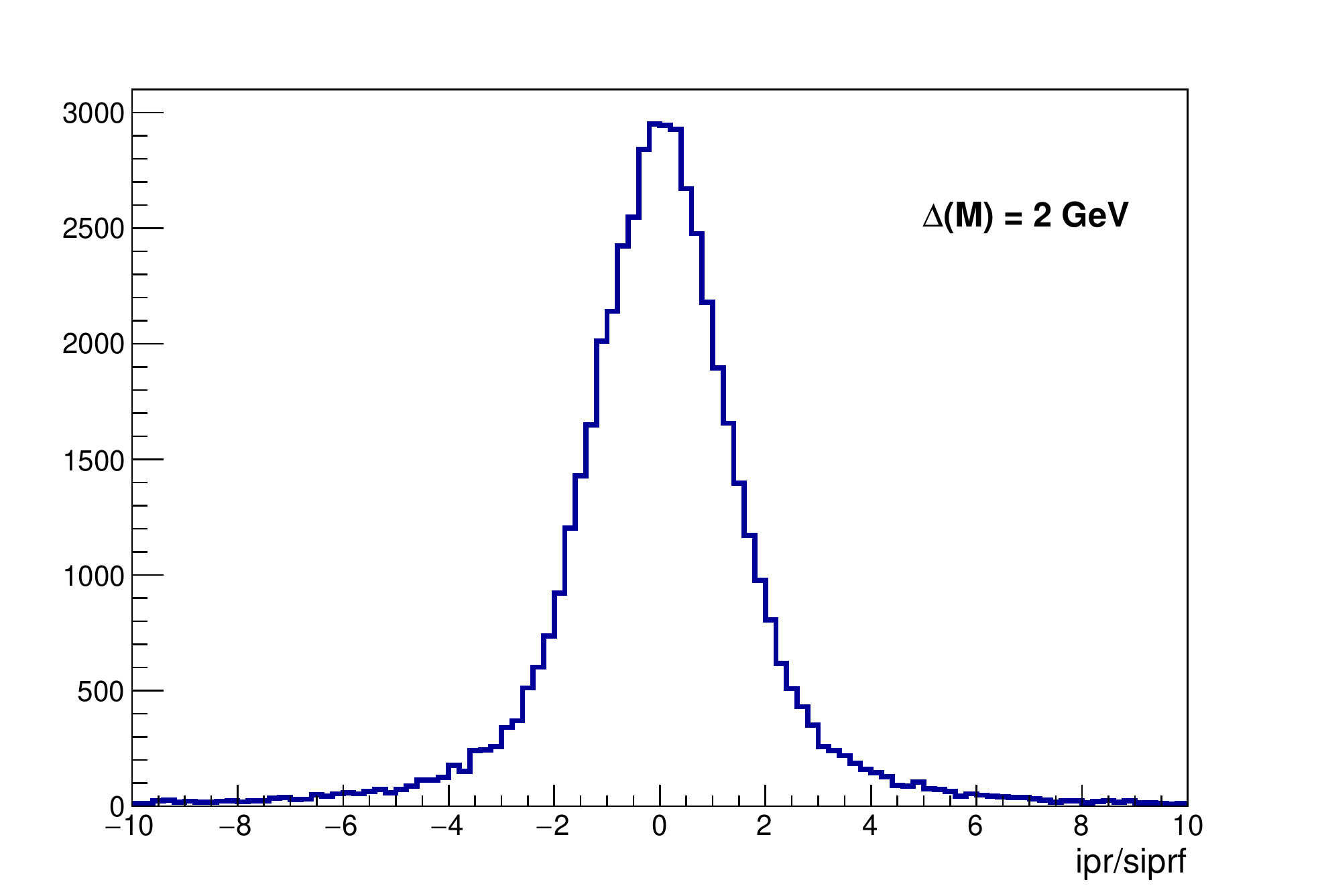}}
    \subcaptionbox{$\Delta(M)$ = 10 GeV}{\includegraphics [width=0.45\textwidth]{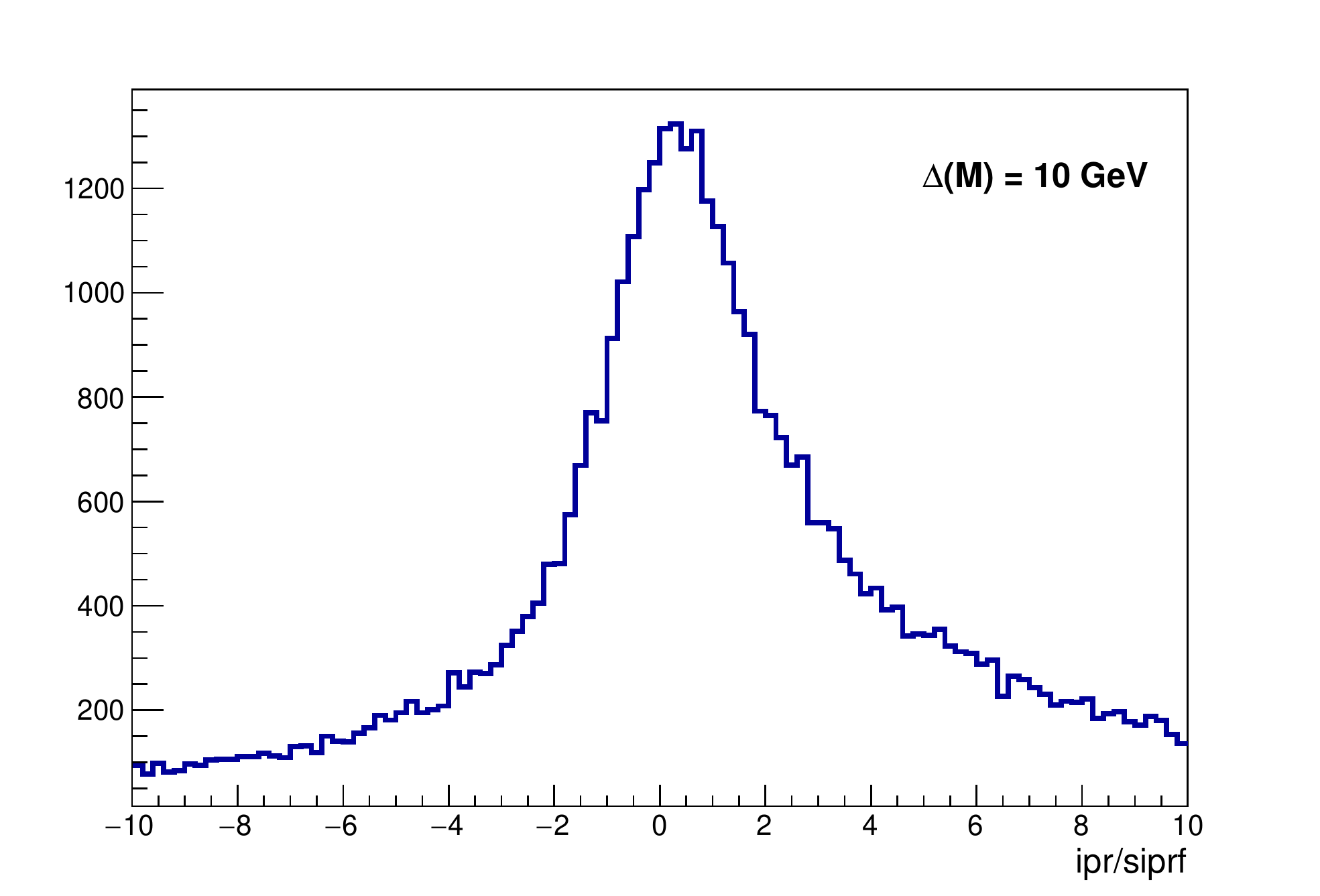}}
    \subcaptionbox{$\Delta(M)$ = 34 GeV}{\includegraphics [width=0.45\textwidth]{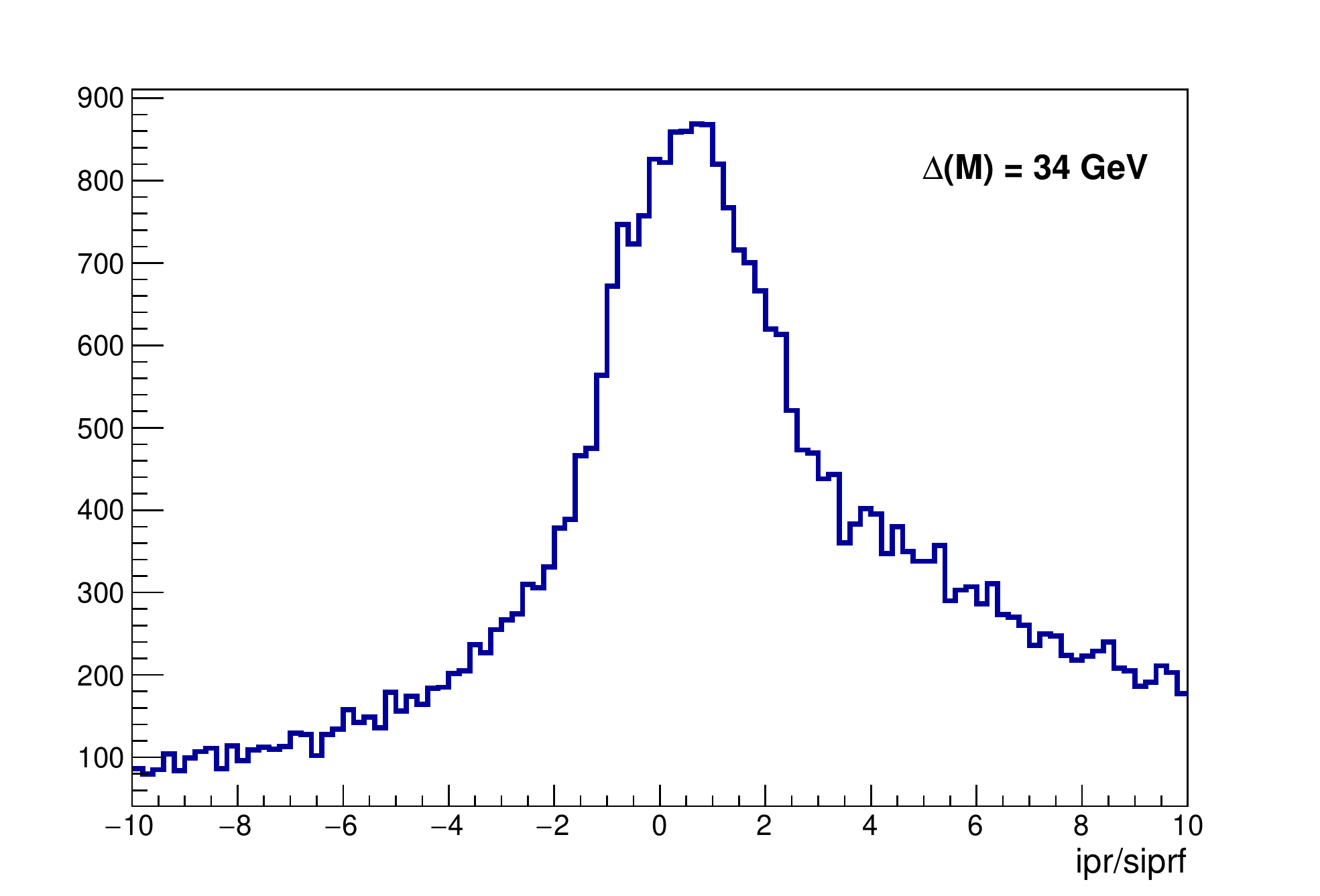}}
    \caption{Transverse impact parameter significance of tracks from overlay and from signals with $M_{\widetilde{\tau}}$ =
    240 GeV, and
    different mass-differences.}
    \label{ip_all}
  \end{figure}

  In the type of event-topologies the signal gives rise to, {\it viz.} two displaced $\tau$ decays, no primary vertex
  can be fitted. However, at ILC the transverse beam-spot size is significantly smaller than the measurement
precision, so that in that plane, the primary vertex position is known, even without fitting.
The longitudinal size of the beam-spot, on the other hand, is much larger than the measurement
precision. 
Figure~\ref{ip_all} shows the  significance of the impact parameter with respect to the origin in the transverse 
direction 
for overlay and signal tracks for different mass differences.
Here, the impact parameter has been {\it life-time signed}, i.e. the sign is determined by the position w.r.t. to the
beam-spot where the
track intersects the axis of the jet it belongs to, taking the direction of the jet as the positive direction.
For the overlay tracks, the expected  insignificant transverse impact parameters for tracks originating inside the beam-spot 
is observed.
For the signal case, one observes more and more significant deviation in the impact parameters, reflecting that the tracks
do {\it not} originate from inside the beam-spot.  
Hence, for model points with mass differences $\le$ 10 GeV, we select tracks with $|cos{\theta}| < 0.7$, 
and the transverse impact parameter significance larger than 2.
If the mass difference is $>$ 10 GeV,
we instead require that.
$P_T > 2$~GeV.

\begin{figure}[htbp]
    \centering
    \subcaptionbox{$\Delta(M) = $ 3 GeV}{\includegraphics [width=0.45\textwidth]{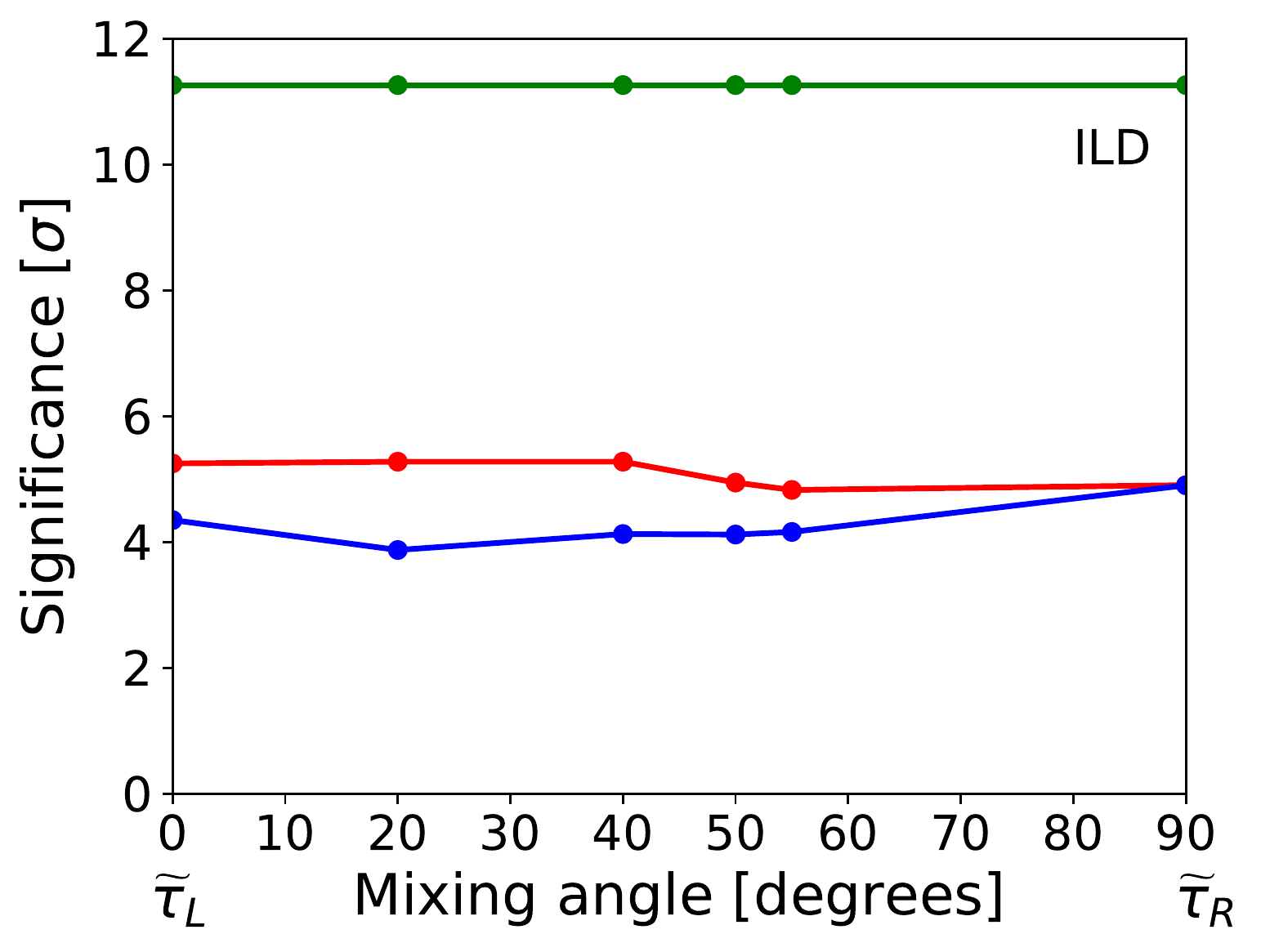}}
    \subcaptionbox{$\Delta(M) = $ 10 GeV}{\includegraphics [width=0.45\textwidth]{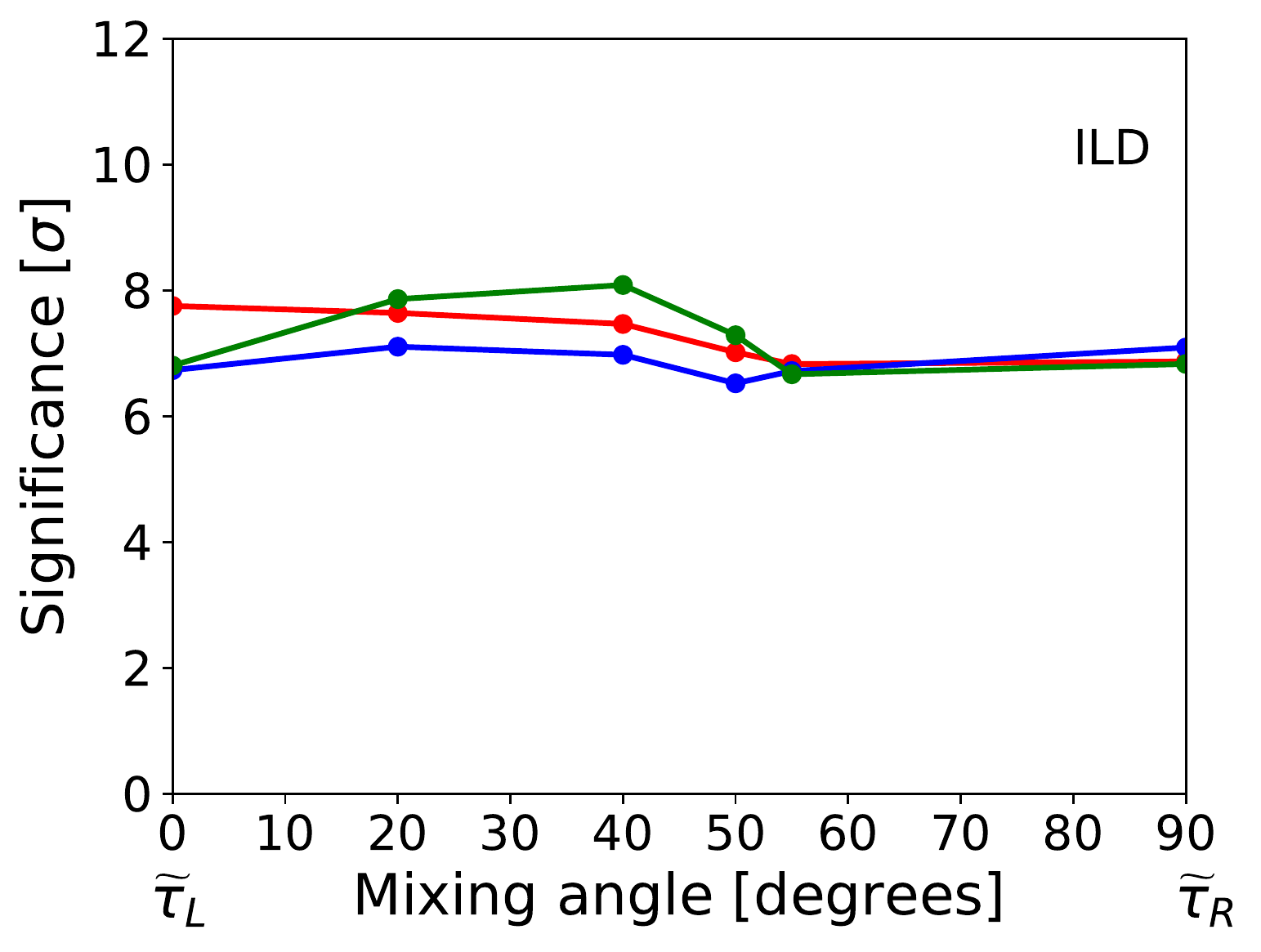}}
    \vspace*{0.4cm}
    \caption{Number of sigmas for a $\widetilde{\tau}$ with mass 240 GeV and different mass-differences.
      The plot assumes H20 ILC scenario combining both polarisations using the likelihood-ratio
      statistic. Blue lines correspond to the case with all the tracks and the red ones after rejecting tracks
      not satisfying the cuts described in the text. The green curves correspond to the study without overlay
      tracks. 
      }
    \label{nb_sigmas_fullsimu_sgv}
  \end{figure}

While these cuts clearly occasionally will remove tracks from the $\tau$ decay resulting in
the event being lost as a candidate (since the strict topology requirements of exactly two $\tau$-jets
with the correct charges will not be met), this is compensated by previous lost candidates being salvaged
by removing wrongly included tracks from overlay in the topology evaluation.
This, together with the fact that many background events only became candidates because of the presence of
overlay tracks, and so will be discarded by the cuts, results in a net amelioration of
the sensitivity.
 Figure~\ref{nb_sigmas_fullsimu_sgv} shows the results obtained with and without cuts together with the results
 obtained if the overlay tracks were added neither to signal nor background.
For the case with the smallest mass difference,
shown in Figure~\ref{nb_sigmas_fullsimu_sgv}(a), 
there is a strong reduction of the significance when adding overlay tracks.
For the larger mass-difference (Figure~\ref{nb_sigmas_fullsimu_sgv}(b)), there is only a slight degradation
due to the overlay tracks. For both mass differences, the overlay removal procedure indeed ameliorates the sensitivity. 

\section{Exclusion/discovery limits\label{sec:limits}}
 
\begin{figure}[htbp]
    \centering
    \includegraphics [width=0.5\textwidth]{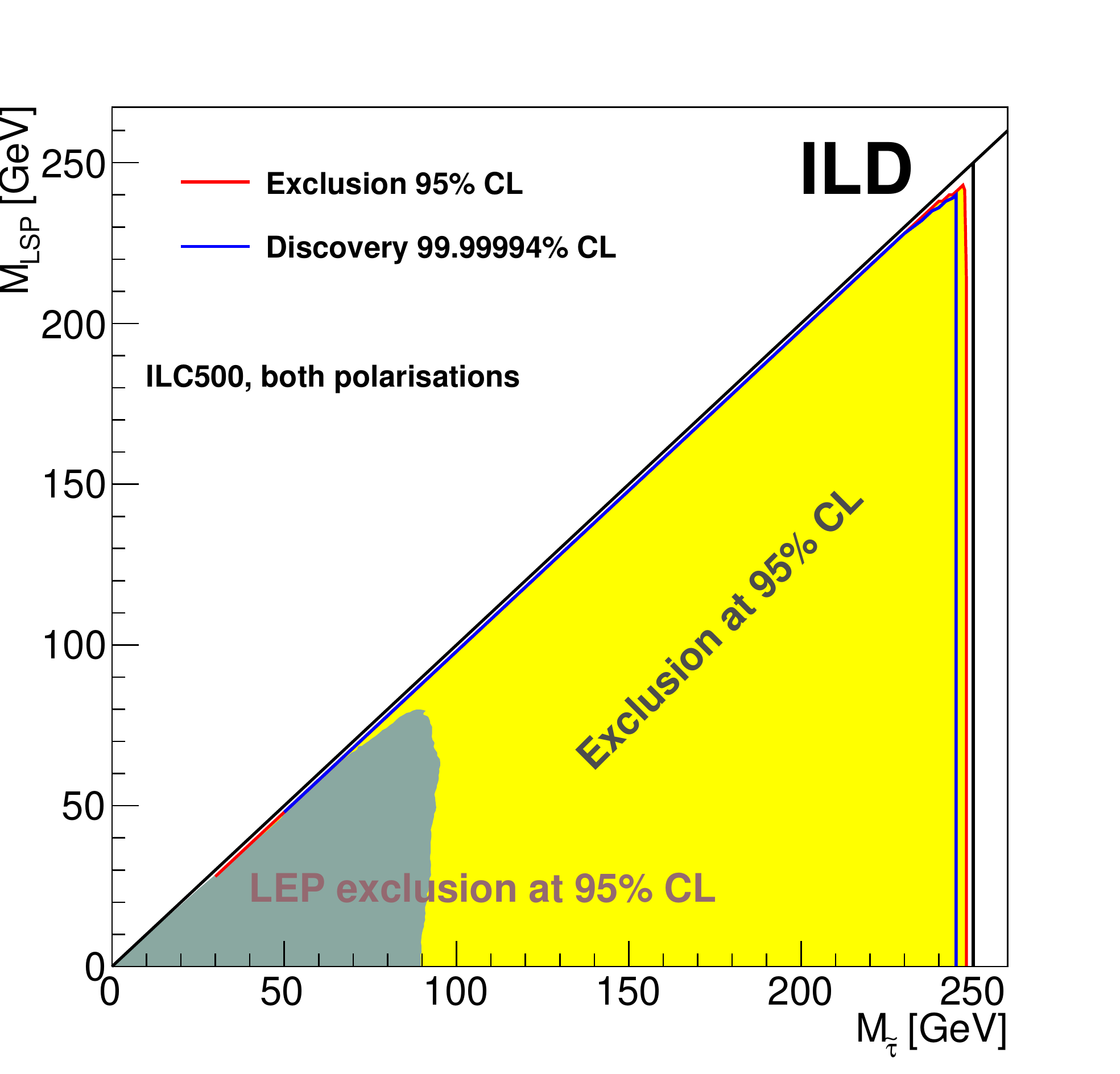}
    \caption{
      Exclusion and discovery $\widetilde{\tau}$ limits from the current studies compared to the ones from LEP studies~\cite{LEPSUSYWG/04-01.1}.
      The ILC region corresponds to the full H20 data-set at $E_{CMS}$ = 500 GeV, i.e. 1.6 ab$^{-1}$ at each of the beam polarisations
      $\mathcal{P}_{+-}$ and  $\mathcal{P}_{-+}$.}
    \label{exclusion_mstauvsmneu}
  \end{figure}

 The exclusion and discovery limits extracted from this study are shown in Figure~\ref{exclusion_mstauvsmneu}.
    They assume the lightest $\widetilde{\tau}$ to be the NLSP and the lightest neutralino the LSP,
    and are valid for any  $\widetilde{\tau}$ mixing angle.
  It is also relevant to compare these results with the current $\widetilde{\tau}$ limits coming from
  LEP. The LEP limits are also valid for any value of the not shown
  model parameters.

\begin{figure}[htbp]
    \centering
    \includegraphics [width=0.5\textwidth]{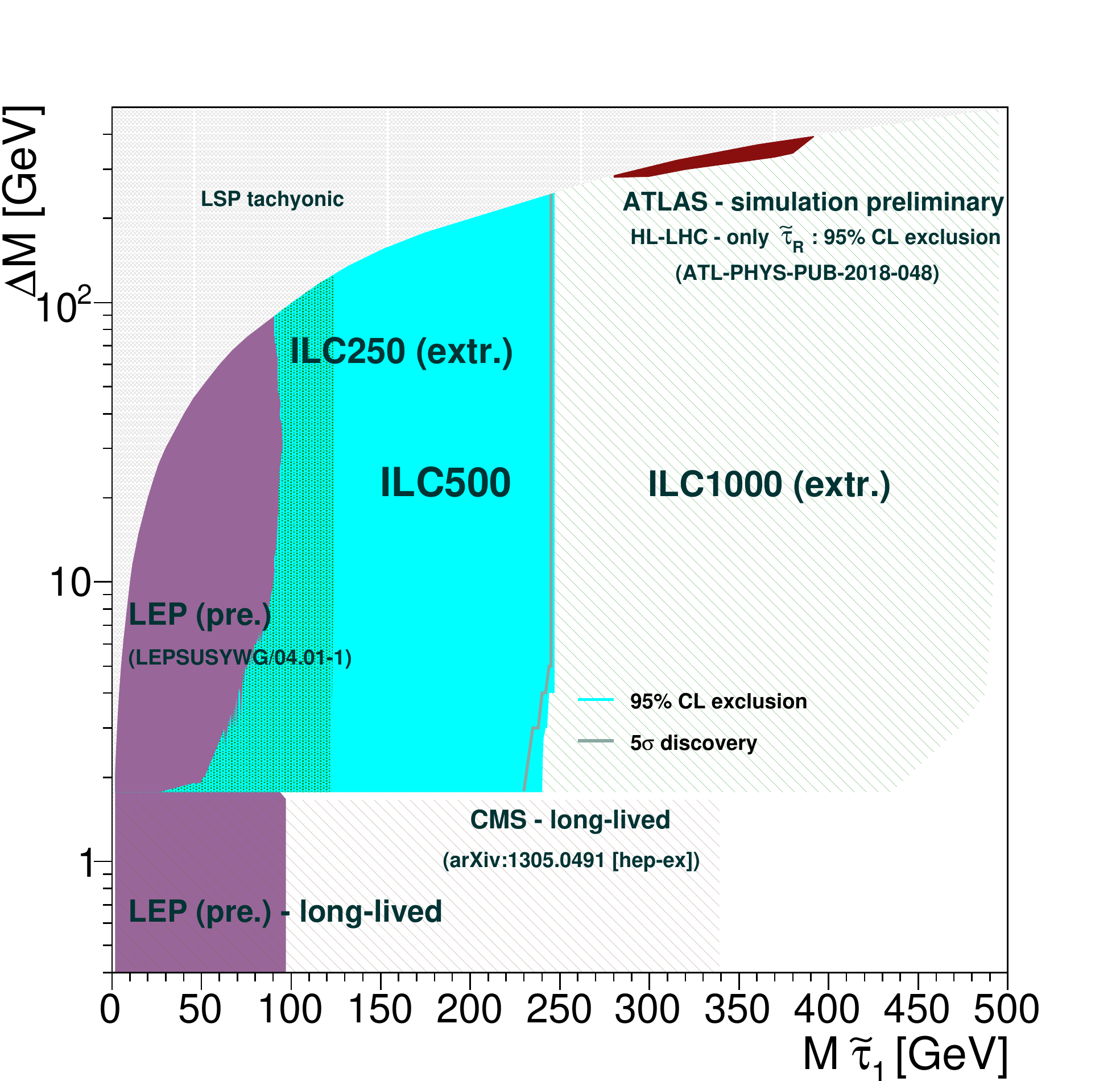}
    \caption{$\widetilde{\tau}$ limits in the $M_{\widetilde{\tau}}$-$\Delta$M plane. ILC results from the current studies are shown together with limits
      from LEP~\cite{LEPSUSYWG/04-01.1} and projection for HL-LHC~\cite{ATLAS:2018diz}. The ILC region corresponds to the full H20 data-set
      at $E_{CMS}$ = 500 GeV, i.e. 1.6 ab$^{-1}$ at each of the beam polarisations $\mathcal{P}_{+-}$ and  $\mathcal{P}_{-+}$.
      The region with mass differences below the mass of the ${\tau}$ is also shown with LEP ~\cite{LEPSUSYWG/02-05.1,LEPSUSYWG/02-09.2} and LHC
      results~\cite{CMS:2013czn}, even if it is not covered by this study. In addition, the extrapolation of
      the ILC current results to the ILC 250~\,GeV and 1~\,TeV running scenarios is shown.}
    \label{exclusion_mstauvsdm}
  \end{figure}

  The projection of the limits in the $M_{\widetilde{\tau}}$-$\Delta$M plane is shown in figure~\ref{exclusion_mstauvsdm}. The
  region for mass differences below the mass of the $\tau$, not included in the current study, is shown for
  completeness. In the region with $\Delta$M larger than $M_{\tau}$ exclusion and discovery ILC limits
  are compared to the ones from LEP.
  The projected HL-LHC limits are also shown. 
  Since they are highly model-dependent, the comparison in
  this case have to be taken with care: here limits considering only the $\widetilde{\tau}_{R}$-pair production
  are shown, since, while still being optimistic, they are closest to the ones expected for the lightest
  $\widetilde{\tau}$ at minimal cross-section.
  It should be noted that the HL-LHC projection is only an exclusion limit - no discovery potential is expected.

  For the region with $\Delta$M smaller than $M_{\tau}$ results from LEP and LHC are shown.
  The LEP studies cover not only the region where the $\widetilde{\tau}_{1}$ travels through the
  detector without decaying but also the region with decays at a certain distance from the production
  vertex. In those regions acoplanar leptons, tracks with large impact parameters and kinked tracks
  are looked for, depending on the $\widetilde{\tau}_{1}$ lifetime~\cite{LEPSUSYWG/02-05.1,LEPSUSYWG/02-09.2}.
The figure also shows the extrapolation of the ILC limits for the scenarios
  with centre-of-mass energy 250~\,GeV and 1~\,TeV.

\subsection{Comparison with previous study\label{sec:newold}}
\begin{figure}[htbp]
    \centering
    \includegraphics [width=0.7\textwidth]{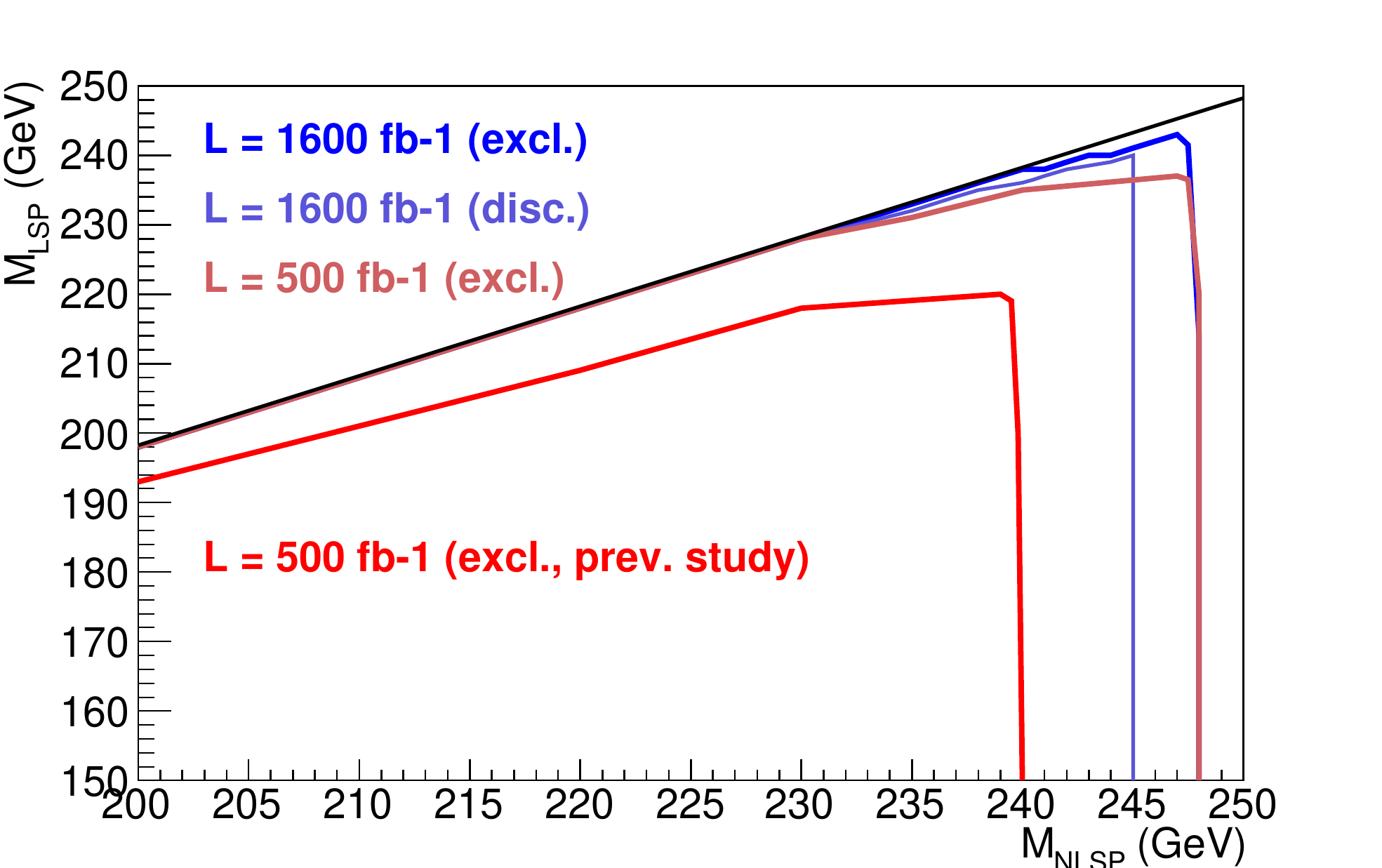}
    \caption{ Exclusion and discovery $\widetilde{\tau}$ limits as a function of the $\widetilde{\tau}$ and LSP masses.
      Exclusion limits from previous ILC studies~\cite{Berggren:2013vna} are also shown, as well as  the current
     analysis applied to the same data-sample as the previous one, i.e. only the  $\mathcal{P}_{+-}$ beam polarisation sample.}
    \label{exclusion_mstauvsmneu_prev}
  \end{figure}

    Results from previous ILC studies,
  which used a $\int \mathcal{L} $dt = 500~\,fb$^{-1}$ sample of  events with beam polarisation  $\mathcal{P}_{+-}$
  are shown in Figure ~\ref{exclusion_mstauvsmneu_prev}, together with 
  results obtained under those conditions with the present analysis.
  The
  comparison of these two curves shows that the extension of the limits is not only due to an increase of the
  total integrated luminosity but also to an improvement of the analysis. The main reason of this improvement
  is the application of individual limits depending on the $\widetilde{\tau}$ mass and the mass difference.
  The previous studies were only making a difference for mass differences above or below 10~\,GeV and
  were not optimised for the low mass difference region. Another difference in the analysis is
  a change in the ${\tau}$-identification algorithm, excluding events with two jets consisting of single leptons of
  the same flavour.

\section{Outlook and conclusions\label{sec:concl}}
The capability of the ILC for excluding/discovering $\widetilde{\tau}$-pair production up to a
few GeV below the kinematic limit, without model dependencies and even in the worst scenario,
has been shown.

The worst scenario for $\widetilde{\tau}$-pair production at the ILC was reviewed taking into account
ILC beam polarisation conditions. Equal sharing of $P(e^{-},e^{+})=(+80\%,-30\%)$ and $P(e^{-},e^{+})=(-80\%,+30\%)$
foreseen in H20 ensures a quite uniform sensitivity to all mixing angles.

The effect of overlay tracks on the signal/background ratio for $\widetilde{\tau}$ searches was analysed.
It was found that the effect is considerable, both for small and moderate mass differences.
Cuts to mitigate the effect were studied and applied.  

The calculation of the exclusion/discovery limits in the region with mass differences below the
${\tau}$ mass, meaning an exponential increase of the $\widetilde{\tau}$ lifetime and consequently
a study of long-lived particles going through or decaying in different parts of the detector, 
is foreseen for a future study, as is the verification that events containing {\it only} overlay tracks
cannot fake signal.


   
\section{Acknowledgements\label{sec:ackn}}
  We would like to thank the LCC generator working group for producing the Monte Carlo samples used in this study.
  We also thankfully acknowledge the support by the the Deutsche Forschungsgemeinschaft (DFG, German Research Foundation) under 
  Germany's Excellence Strategy EXC 2121 ``Quantum Universe'' 390833306.
  This work has benefited from computing services provided by the German National Analysis Facility (NAF)~\cite{Haupt:2010zz}.

\printbibliography

\end{document}